\documentclass[aps,prl,notitlepage,reprint,twocolumn]{revtex4-2}
\usepackage[utf8]{inputenc}
\usepackage[T1]{fontenc}
\usepackage{epsfig}
\usepackage{epstopdf}
\usepackage{epsfig}
\usepackage{dcolumn}
\usepackage{color}
\usepackage{amsmath}
\usepackage{amssymb}
\usepackage{graphicx}
\usepackage{hyperref}
\usepackage{nameref}
\usepackage{indentfirst}
\usepackage{color}
\usepackage{changes}

\usepackage{bm}
\usepackage{marvosym}
\usepackage{setspace}
\usepackage{pifont}

\usepackage{lineno}

\begin{document}
\title{Nonsmooth Bloch Oscillations in Non-Hermitian Systems}

\author{Yanyan He$^1$}\email{he.yanyan.c6@tohoku.ac.jp}

\author{Tomoki Ozawa$^{1,2}$}
\email{tomoki.ozawa.d8@tohoku.ac.jp}

\affiliation{$^1$Advanced Institute for Materials Research (WPI-AIMR), Tohoku University, Sendai 980-8577, Japan\\
$^2$RIKEN Center for Interdisciplinary Theoretical and Mathematical
Sciences (iTHEMS), RIKEN, Wako, Saitama 351-0198, Japan} 


\begin{abstract}
Bloch oscillations (BOs) represent a paradigmatic manifestation of force-driven band dynamics in lattices, conventionally characterized by continuous quasimomentum evolution and smooth periodic trajectories. Here we show that non-Hermiticity qualitatively changes this paradigm, introducing nonsmooth BOs characterized by cusp-like real-space trajectories and abrupt momentum jumps. Reversing the force switches the dynamics between smooth and nonsmooth BOs, enabling strongly nonreciprocal dynamics. We derive a universal relation for the critical force that governs the nonsmooth-to-smooth dynamical transition in general one-dimensional lattices, which is induced by a pitchfork bifurcation in momentum space. Our work establishes a unified framework for non-Hermitian BOs and uncovers a fundamental bifurcation mechanism that can generate nonsmooth wave transport in entirely linear systems.
\end{abstract}

\date{\today}
\maketitle

\textit{\textcolor{blue}{Introduction}}---Coherent wave dynamics under external forces provides a fundamental probe of quantum transport in periodic lattices. A paradigmatic manifestation is the 
Bloch oscillation (BO) \cite{bloch1929quantenmechanik,zener1934theory}, in which a wave packet undergoes periodic and smooth motion driven by a constant force. Its conventional semiclassical picture is governed by the acceleration theorem: the external force drives the quasimomentum to evolve continuously and linearly in time, while the band group velocity converts this evolution into a smooth periodic trajectory \cite{hartmann2004dynamics}. BOs have been observed in a broad range of Hermitian real-space platforms \cite{esaki1970superlattice,leo1992observation,feldmann1992optical,waschke1993coherent,dahan1996bloch,wilkinson1996observation,ferrari2006long,gustavsson2008control,morandotti1999experimental,pertsch1999optical,chiodo2006imaging,sapienza2003optical,sanchis2007acoustic} and synthetic lattices \cite{chen2021real,oliver2023bloch,xie2026quantum}.

Non-Hermiticity fundamentally enriches wave dynamics by introducing gain, loss, and nonreciprocal couplings, giving rise to exotic features without Hermitian counterparts \cite{bender2007making,feng2017non,el2018non,ashida2020non,wang2021topological,bergholtz2021exceptional,ding2022non,okuma2023non,hatano1996localization,yao2018edge}. Such systems are typically described by effective non-Hermitian (NH) Hamiltonians with complex band dispersions \cite{gong2018topological,xiao2025non}. In particular, wave dynamics is governed not only by the real part of the band, but also by its imaginary part through momentum-dependent amplification and attenuation \cite{muschietti1993real,silberstein2020berry,solnyshkov2021quantum,longhi2022non,hu2023wave,dong2026non,tao2026imaginary}. Consequently, such complex band dispersions can induce anomalous transport phenomena, including robust one-way wave propagation \cite{longhi2015robust}, self-acceleration of wave packets \cite{longhi2022non,xue2024self,tao2026imaginary}, dynamic skin effects \cite{li2022dynamic,guo2022theoretical,li2024observation}, and NH wave-packet jumps \cite{he2026anomalous,beck2025wave,jana2026solution}. Under an external force, a wave packet therefore experiences both force-driven quasimomentum drift and non-Hermiticity-induced momentum selection. Although several aspects of BOs in NH systems have been reported~ \cite{efremidis2004bloch,longhi2009bloch,regensburger2012parity,longhi2014exceptional,longhi2015bloch,wimmer2015observation,xu2016experimental,graefe2016quasiclassical,gong2018topological,qin2020discrete,peng2022manipulating,zhang2024dynamics,zhang2024extended,li2024fate,chakrabarty2025fate,qin2024geometry,fahara2025exact}, a unified theoretical framework for NH BO dynamics is still missing, and whether non-Hermiticity can qualitatively alter the canonical smooth BOs remains an open question.

In this Letter, we establish a general framework for NH BOs and demonstrate that non-Hermiticity induces a momentum-space pitchfork bifurcation, leading to the breakdown of conventional smooth BOs. Due to the interplay between the force-driven and non-Hermiticity-induced momentum evolution, the momentum extrema exhibit multiple branches, where the dominant momentum switches from one branch to another. Such a sudden switching of the dominant momentum induces nonsmooth BOs featuring cusp-like trajectories and abrupt changes in group velocities. Since the bifurcation depends on the force direction, reversing the force switches the dynamics between nonsmooth and smooth BOs. Furthermore, we reveal a nonsmooth-to-smooth dynamical transition at a critical force strength, i.e., the bifurcation point, where these multiple branches coalesce into a single branch, thereby restoring the smooth BOs. 
We demonstrate this phenomenon by studying the wave-packet dynamics driven by a constant force in general one-dimensional NH lattices and derive an analytical universal relation for the critical force, which forms a \textit{bifurcation surface} in the parameter space that governs the transition. 

\begin{figure}[ht!]
\centering
\includegraphics[width=\linewidth]{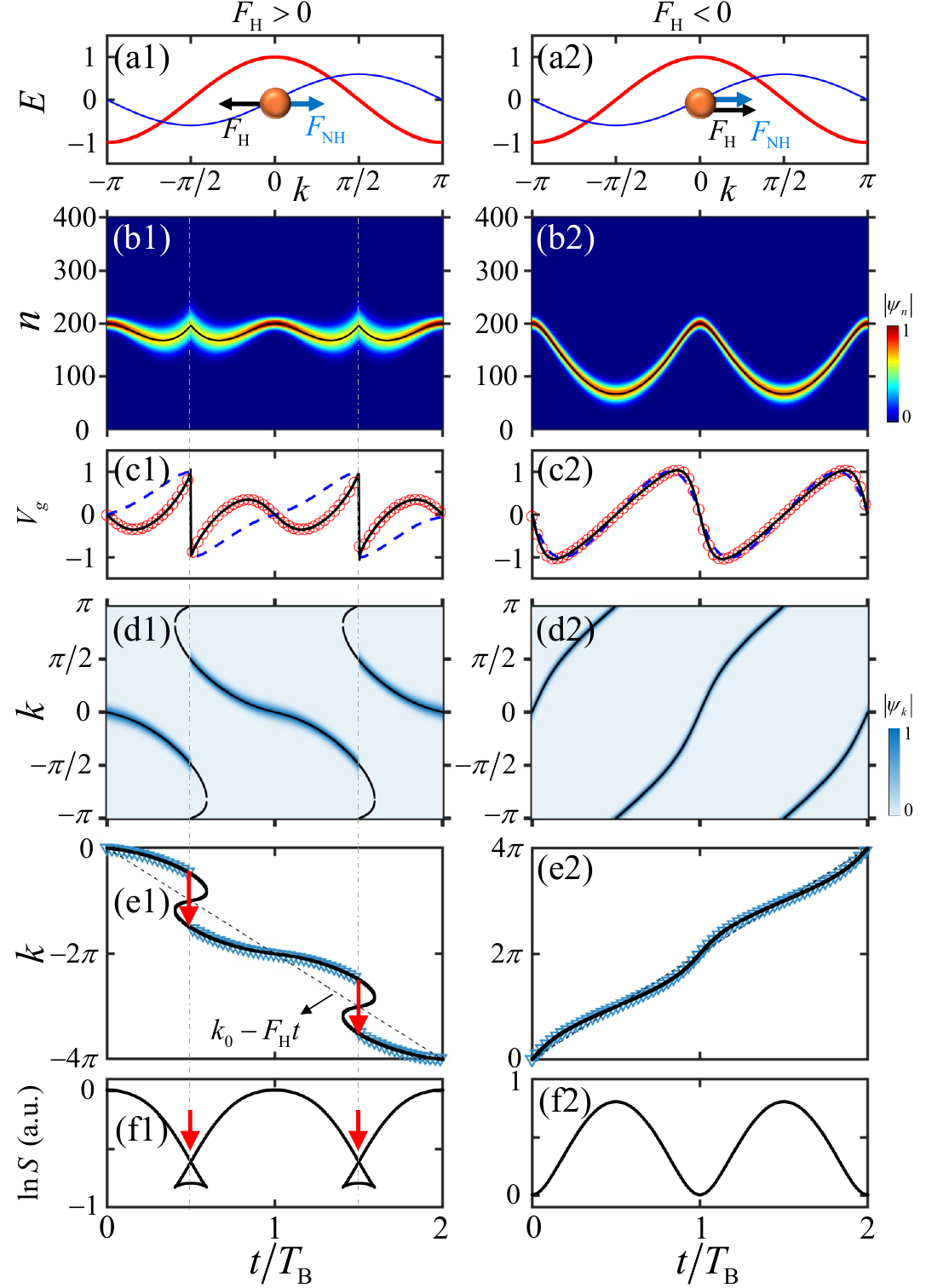}
\caption{(a1,a2) Band dispersions, $E_R(k)$ (red) and $E_I(k)$ (blue). (b1,b2) Real-space wave-packet dynamics, where the black solid lines denote the predicted center of mass trajectory $\langle n(t)\rangle$ obtained from Eq.~(\ref{n_average}). (c1,c2) Numerical (red circles) and the predicted group velocity $V_g(t)$ from Eq.~(\ref{V}) (black solid lines), and the Hermitian counterpart $\langle dE_R/dk\rangle$ (blue dashed lines). (d1,d2) Momentum-space dynamics, where the solid lines denote $k_{\text{e}}(t)$. (e1,e2) Same as (d1,d2), but plotted in multiple Brillouin zones. The blue triangles represent the numerically extracted $k_{\text{m}}(t)$. (f1,f2) Amplitude of the momentum-space wavefunction evaluated at $k_{\text{e}}(t)$, i.e., $|S(k_{\text{e}},t)|$.~a.u.,~arbitrary units.~Parameters for (a1-f1) are $\mathcal{J}=0.5,~\mathcal{J}'=0.3,~\theta=0,~\theta'=\pi/2,~\sigma=5,~k_0=0,N=400,~n_0=N/2$, and $F_{\text{H}}=0.015$. Parameters for (a2-f2) are identical except $F_{\text{H}}=-0.015$.}
\label{fig1}
\end{figure}

\textit{\textcolor{blue}{Nonreciprocal nonsmooth BOs}}--- {We first numerically demonstrate the phenomenon of nonreciprocal nonsmooth BOs considering} the wave-packet dynamics in a {one-dimensional} NH lattice {with nearest-neighbor hoppings} driven by a constant force $F_{\text{H}}$, which is described by {the following Hamiltonian:}
\begin{align}
    H=\sum_n J_L|n\rangle\langle n+1|+J_R|n+1\rangle\langle n|+F_{\text{H}}n|n\rangle \langle n|. \label{H}
\end{align}
Here $|n\rangle$ denotes the 
state localized at the lattice site $n$, and $J_{L,R}\in \mathbb{C}$ denote the {generally complex} left/right hopping amplitudes. The corresponding band dispersion is $E(k)\equiv E_R (k) + i E_I (k)$, where $E_{R(I)}$ denotes the real (imaginary) part of the energy dispersion and can be written as $E_R (k) = 2\mathcal{J}\cos(k-\theta)$ and $E_I(k) = 2\mathcal{J}'\cos(k-\theta')$. {Here $\mathcal{J}(\ge0)$, $\mathcal{J}'(\ge0)$, $\theta$, and $\theta'$ are functions of $J_L$ and $J_R$} [Supplementary Material (SM) \cite{supp}]. Without loss of generality, we set $\theta=0$ in the analysis below. 

We first consider the 
{case} with $\theta'=\pi/2$, which has nonreciprocal couplings $J_{L(R)}=\mathcal{J}\pm\mathcal{J}'$ and band dispersions $E_R (k) = 2\mathcal{J}\cos(k)$ and $E_I(k) = 2\mathcal{J}'\sin(k)$ [Figs.~\ref{fig1}(a1) and \ref{fig1}(a2)]. {We consider time evolution of the wavefunction which is initially a Gaussian wave packet} $\psi_n(0)=\frac{1}{(2\pi\sigma^2)^{1/4}}e^{-\frac{(n-n_0)^2}{4\sigma^2}}e^{ik_0(n-n_0)}$, where $k_0,n_0$ denote the initial momentum and position, respectively. We show the wave dynamics for $F_{\text{H}}=\pm0.015$ by plotting the numerically calculated normalized wavefunction \cite{longhi2015robust,longhi2015bloch} over two Bloch periods with $T_\text{B}=2\pi/|F_{\text{H}}|$, as shown in Figs.~\ref{fig1}(b1) and~\ref{fig1}(b2), respectively. We see the BOs exhibit strongly nonreciprocal features for opposite forces. For positive $F_{\text{H}}$, the wave packet undergoes a nonsmooth oscillatory evolution {with a cusp-like trajectory} accompanied by abrupt changes in the group velocity [red circles in Figs.~\ref{fig1}(c1)], indicating the breakdown of the conventional smooth BOs. Meanwhile, the corresponding momentum-space normalized wavefunction $\psi_k(t)$ exhibits periodic discontinuous jumps [Fig.~\ref{fig1}(d1)] (We note that, depending on parameters, there can also be jumps in real space \cite{supp}). In contrast, for negative $F_{\text{H}}$, the wave dynamics in real space remains {smooth} [Fig.~\ref{fig1}(b2) and~\ref{fig1}(c2)], and the momentum-space wavefunction evolves continuously [Fig.~\ref{fig1}(d2)], similar to conventional BOs in Hermitian systems.

\textit{\textcolor{blue}{Universal theoretical description of nonreciprocal nonsmooth BOs}}--- 
To uncover the origin of this phenomena, we {now consider general single-band one-dimensional NH lattices with possibly longer-range hoppings. We denote the complex band dispersion by $E(k)$. We consider an initial Gaussian wave packet, which in momentum space is} $S(k,0)=Ce^{-\sigma^2(k-k_0)^2}e^{-ikn_0}$ with $C=\left(2\sigma^2/\pi\right)^{{1}/{4}}$. Its time evolution is governed by \cite{longhi2015bloch,hartmann2004dynamics}
\begin{align}
i\partial_tS(k,t)=E(k)S(k,t)+iF_{\text{H}}\partial_kS(k,t).\label{S_partial}
\end{align}
The general solution is $
    S(k,t)=S(k+F_{\text{H}}t,0)e^{-i\Phi(k,t)}$, where $\Phi(k,t)=\Phi_R(k,t)+i\Phi_I(k,t)$ with 
$\Phi_{R(I)}=\int_0^tdt' E_{R(I)}(k+F_{\text{H}}t-F_{\text{H}}t')$.
The amplitude of $S(k,t)$ is
\begin{align}
    |S(k,t)|=Ce^{-\sigma^2(k+F_{\text{H}}t-k_0)^2}e^{\Phi_I(k,t)}\equiv Ce^{M(k,t)},\label{S_abs}
\end{align}
with $M(k,t)=-\sigma^2(k+F_{\text{H}}t-k_0)^2+\Phi_I(k,t)$, which shows that the {amplitude evolution in momentum space} is governed by two competing contributions. The first originates from the external constant force $F_\text{H}$, which drives a linear drift of the wave-packet momentum {toward $-F_{\text{H}}$ direction in momentum space} \cite{hartmann2004dynamics}, as indicated by the black arrows in Figs.~\ref{fig1}(a1) and~\ref{fig1}(a2). The other arises from the NH component, {$\Phi_I(k,t)$}, which drives the momentum toward the direction with increasing gain, {${dE_I(k)}/{dk}$}
~\cite{muschietti1993real,silberstein2020berry,he2026anomalous,tao2026imaginary}, as denoted by the blue arrows in Figs.~\ref{fig1}(a1) and~\ref{fig1}(a2) {(see also SM~\cite{supp})}. We therefore refer to this contribution as the { effective NH force} $F_\text{NH}$. {Since we have both contributions, the extremal {momentum} $k_{\text{e}}(t)$, at which $|S(k,t)|$ reaches an extremum, is determined by their interplay, and is given by the condition ${\partial_k M(k,t)}=0$. We thus obtain an implicit equation which determines $k_{\text{e}}(t)$ (see SM~\cite{supp}) }
\begin{align}
\frac{E_I(k+F_{\text{H}}t)-E_I(k)}{2\sigma^2 F_{\text{H}}}=k+F_{\text{H}}t-k_0,\label{partial_M}
\end{align}
which is the NH generalization of the acceleration theorem for the extremal momentum in NH systems. In the Hermitian limit ($E_I=0$), this relation recovers the conventional acceleration theorem, $k_{\text{e}}(t)=k_0-F_{\text{H}}t$ \cite{hartmann2004dynamics}.

In real space, the center of mass and the corresponding group velocity of the wave packet are given by (see SM \cite{supp})
\begin{align}
    \langle n(t) \rangle&=\frac{\sum_nn|\psi_n(t)|^2}{\sum_n|\psi_n(t)|^2}=n_0+\left\langle\frac{E_R(k+F_{\text{H}}t)-E_R(k)}{F_{\text{H}}}\right\rangle,\label{n_average}\\
    V_g(t)&=\frac{d\langle n(t) \rangle}{dt}=\left\langle \frac{d E_R (k)}{d k}\right\rangle+V_A(t),\label{V}
\end{align}
where 
\begin{align}
V_A(t)&=2\left\langle\frac{E_R(k+F_{\text{H}}t)-E_R(k)}{F_{\text{H}}}E_I(k)\right\rangle \nonumber\\
&-2\left\langle\frac{E_R(k+F_{\text{H}}t)-E_R(k)}{F_{\text{H}}}\right\rangle\langle E_I(k)\rangle. \label{V_A}
\end{align}
The momentum-space average of a function $f(k)$ is defined as $\langle f(k)\rangle=\int dk |S(k,t)|^2f(k)/\int dk |S(k,t)|^2$, which hence implies that the discontinuous evolution in momentum space leads to nonsmooth dynamics in real space. The analytical predictions for $\langle n(t) \rangle$ [black solid lines in Figs.~\ref{fig1}(b1) and~\ref{fig1}(b2)] and $V_g(t)$ [black solid lines in Figs.~\ref{fig1}(c1) and~\ref{fig1}(c2)] are in agreement with the numerical results. The term $V_A(t)$ is the \textit{anomalous group velocity} \cite{he2026anomalous} in the existence of a force, which makes the group velocity different from the Hermitian counterpart, $\left\langle {d E_R}/{d k}\right\rangle$, as shown by the blue dashed lines in Figs.~\ref{fig1}(c1) and~\ref{fig1}(c2). Eqs.~(\ref{partial_M}-\ref{V_A}) constitute a unified framework for NH BO dynamics.

{We can now use this universal framework to understand the nonreciprocal nonsmooth BOs found in Fig.~\ref{fig1}.}
Substituting $E_I(k) = 2\mathcal{J}'\cos(k-\theta')$ into Eq.~(\ref{partial_M}), we obtain the following implicit equation that determines the $k_\text{e}(t)$ (see SM \cite{supp}),
\begin{align}
\frac{F_\text{c}}{F_\text{H}}\sin\left(\frac{F_\text{H}t}{2}\right)\sin\left(k+\frac{F_\text{H}t}{2}-\theta'\right)=k_0-k-F_\text{H}t, \label{ke}
\end{align}
where $F_\text{c}={2\mathcal{J}'}/{\sigma^2}$. The prediction of $k_{\text{e}}(t)$, shown as the solid lines in Figs.~\ref{fig1}(d1) and~\ref{fig1}(d2), agrees well with the numerical results. This agreement is further shown in the extended Brillouin zones, where $k_{\text{e}}(t)$ and numerically extracted momentum $k_{\text{m}}(t)$, corresponding to the maximum of $|\psi_k(t)|$, are plotted as the solid lines and triangles, respectively [Figs.~\ref{fig1}(e1)and~\ref{fig1}(e2)]. For $F_{\text{H}}>0$, whose direction is opposite to that of $F_{\text{NH}}$ near the initial momentum $k_0=0$ [see Fig.~\ref{fig1}(a1)], $k_\text{e}(t)$ admits multiple solutions during the time evolution [Fig.~\ref{fig1}(e1)]. Among these solutions, only $k_{\text{m}}(t)$, corresponding to the maximum of $|S(k_{\text{e}},t)|$, can survive and dominate the observed wave dynamics. This can be seen from Fig.~\ref{fig1}(f1). The amplitude $|S(k_\text{e},t)|$ at the initial solution decreases with time, whereas that at another solution increases. Consequently, the latter dominates the evolution, making the initial wave component hidden. The resulting jump of the dominant momentum, as highlighted by the red solid arrows at the jump time $t^*=T_\text{B}/2$, leads to the nonsmooth trajectory with {cusps} in real space. Because of the periodic changes of $|S(k_\text{e},t)|$ at different dominant momenta [Fig.~\ref{fig1}(f1)], the {cusps} occur periodically. In contrast, for $F_{\text{H}}<0$, whose direction is the same as $F_{\text{NH}}$ [Fig.~\ref{fig1}(a2)], $k_\text{e}(t)$ admits only a single solution throughout the evolution [Figs.~\ref{fig1}(d2)-(f2)]. It hence gives rise to {smooth} real-space dynamics without any {cusps or jumps} [see Figs.~\ref{fig1}(b2) and~\ref{fig1}(c2)]. The finite-size effects on the NH BOs are discussed in SM \cite{supp}. 

\begin{figure*}[t!]
\centering
\includegraphics[width=0.9\linewidth]{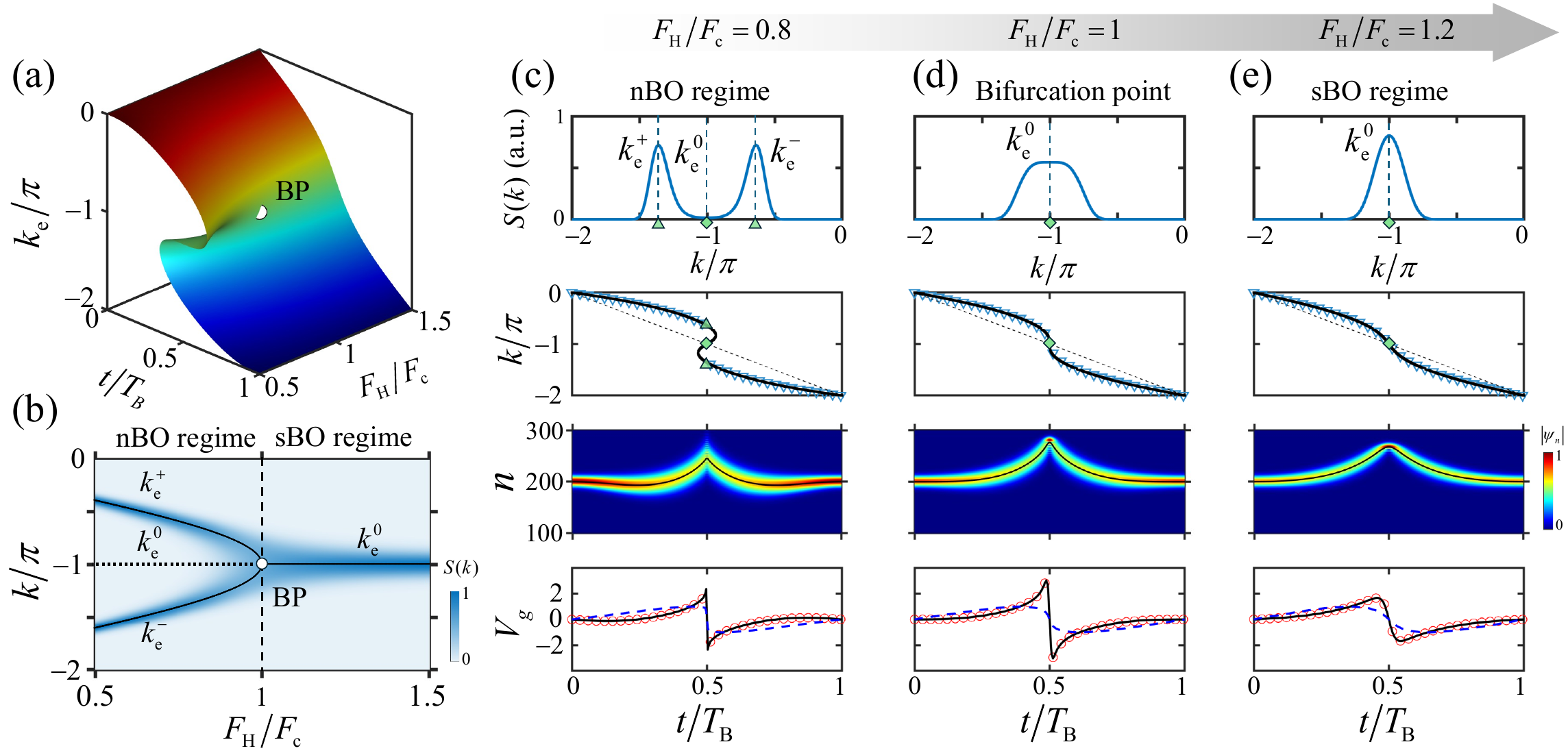}
\caption{(a) Solution of $k_{\text{e}}$ in the $(t,F_{\text{H}})$ plane, where the white dot denotes the bifurcation point (BP). (b) The evolution of $S(k,t=T_\text{B}/2)$ for different $F_{\text{H}}$. The black lines denote the solutions $k_\text{e}(t=T_\text{B}/2)$. Top row of (c)-(e) $S(k,t=T_\text{B}/2)$ for different $F_{\text{H}}$. Second, third, and last row of (c-e) are the same as Figs.~\ref{fig1} (e1), (b1), and (c1) with the same descriptions. The green triangles (diamonds) denote $k_\text{e}^{\pm}$ ($k_\text{e}^{0}$). The parameters are the same as those in Fig.~\ref{fig1} (a1-f1).}
\label{fig2}
\end{figure*}
\textit{\textcolor{blue}{Bifurcation-induced nonsmooth-to-smooth transition}}---
We now analyze the transition between the nonsmooth and smooth BOs. We plot the solutions $k_{\text{e}}(t)$ with respect to $F_{\text{H}}$ in Fig.~\ref{fig2}(a), where we see that, during one oscillation period $T_\text{B}$, $k_{\text{e}}$ can admit three solutions below the critical force $F_{\text{c}}$ highlighted by a white point, while it has only one solution above $F_{\text{c}}$. Since the multiple solutions (single solution) of $k_\text{e}$ induce the nonsmooth (smooth) BOs, there exists a transition from nonsmooth to smooth BOs when increasing $F_\text{H}$. The jump time when the dominant (maximum) momentum switches between different solutions occur at  (see SM \cite{supp}) 
\begin{align}
    t^*=\left(1-\frac{|\theta'-k_0|}{\pi}\right)T_\text{B}. \label{t_star}
\end{align}
For $\theta'=0.5\pi,~k_0=0$ used in above analysis, we obtain $t^*=T_\text{B}/2$, agreeing with the numerical results in Fig.~\ref{fig1}. 
The geneal equation that determines the solution $k_\text{e}(t)$ in Eq.~(\ref{ke}) at $t=T_\text{B}/2$ can be rewritten as (see SM \cite{supp})
\begin{align}
    Y=A\sin(Y),\label{Y}
\end{align}
where $Y=k_\text{e}(t=T_\text{B}/2)+\pi$ and $A={F_\text{c}}/{F_\text{H}}$. 

The {number of solutions} is determined by {the value of} $A$, which induces a pitchfork bifurcation from three solutions to one solution, as shown by the black lines in Fig.~\ref{fig2}(b).  When $A>1$, i.e., $F_{\text{H}}<F_\text{c}$, there are three solutions $Y=0,~\pm Y_0$, where $Y_0$ is determined by ${1}/{A}={\sin(Y_0)}/{Y_0}.$ In this case, $k_\text{e}(t=T_\text{B}/2)$ admits three solutions $\{k_\text{e}^0=-\pi,~ k_\text{e}^{\pm}=k_\text{e}^0\pm Y_0\}$, where $k_\text{e}^0$ and $k_\text{e}^{\pm}$ correspond to the local minimum (green diamond) and local maxima (green triangles) of the momentum-space wavefunction $S(k,t=T_\text{B}/2)$, respectively [see the top row in Fig.~\ref{fig2}(c)]. As time evolves, the global maximum momentum that dominates the evolution jumps between these two peaks (local maxima) [the second row of Fig.~\ref{fig2}(c)], resulting in the nonsmooth BO in real space with abrupt changes in the group velocity [the third and last rows of Fig.~\ref{fig2}(c)]. Therefore, $F_\text{H}<F_\text{c}$ indicates the nonsmooth BO regime.

The transition occurs at $A=1$, i.e, $F_{\text{H}}=F_\text{c}$, which is the bifurcation point where these three solutions coalesce into $k_\text{e}^0=-\pi$ [Fig.~\ref{fig2}(b)]. The two peaks of $S(k,t=T_\text{B}/2)$ merge into a single peak (local maximum), corresponding to the critical case where the corresponding BO is still smooth [Fig.~\ref{fig2}(d)]. 

When $A<1$, i.e., $F_{\text{H}}>F_\text{c}$, the system enters the smooth BO regime [Fig.~\ref{fig2}(b)], as there is only one solution with $Y=0$, which gives $k_\text{e}(t=T_\text{B}/2)=k_\text{e}^0=-\pi$. $S(k,t=T_\text{B}/2)$ has only one peak (local maximum), which remains dominant throughout the evolution. The evolution in momentum space is continuous [second row of Fig.~\ref{fig2}(e)] and the corresponding BO in real space is hence smooth [the third and last rows of Fig.~\ref{fig2}(e)].

\textit{\textcolor{blue}{Transition diagram and bifurcation surface}}---We now show that such a nonsmooth-smooth BOs transition can occur for {general values of} $\theta'$ and $k_0$, and derive the critical force $F_\text{H}^*$, which forms a \textit{bifurcation surface }in the parameter space $(\theta',k_0)$. For {general values of} $(\theta',k_0)$, the extremum equation, Eq.~(\ref{ke}), that determines the solutions $k_\text{e}(t^*)$ can also be written {in the form} of Eq.~(\ref{Y}) with the replacements $Y=k_\text{e}(t=t^*)+F_{\text{H}}t^*-k_0$ and $A=\frac{F_\text{c}}{F_{\text{H}}}\sin(\theta'-k_0)$ (see SM\cite{supp}). Therefore, for $A>1$, there exist three solutions $\{k_\text{e}^0=k_0-F_{\text{H}}t^*, k_\text{e}^{\pm}=k_\text{e}^0\pm Y_0\}$, corresponding to the nonsmooth BO regime, whereas, for $A<1$, there exists one solution $k_\text{e}^0$, indicating the smooth BO regime. The bifurcation point occurs at $A=1$, which gives the critical force
\begin{align}
F_{\text{H}}^*=F_\text{c}\sin(\theta'-k_0).
\end{align}
Considering the sign of $F_{\text{H}}$, the regime for non-smooth BOs is \cite{supp}
\begin{align}
\begin{cases}
    F_{\text{H}}<F_{\text{H}}^*,~~ 
 \text{for}~~~F_{\text{H}}>0, \\  
F_{\text{H}}>F_{\text{H}}^*,~~ 
 \text{for}~~~F_{\text{H}}<0. \label{transition}
\end{cases}
\end{align}
\begin{figure}[t!]
\centering
\includegraphics[width=\linewidth]{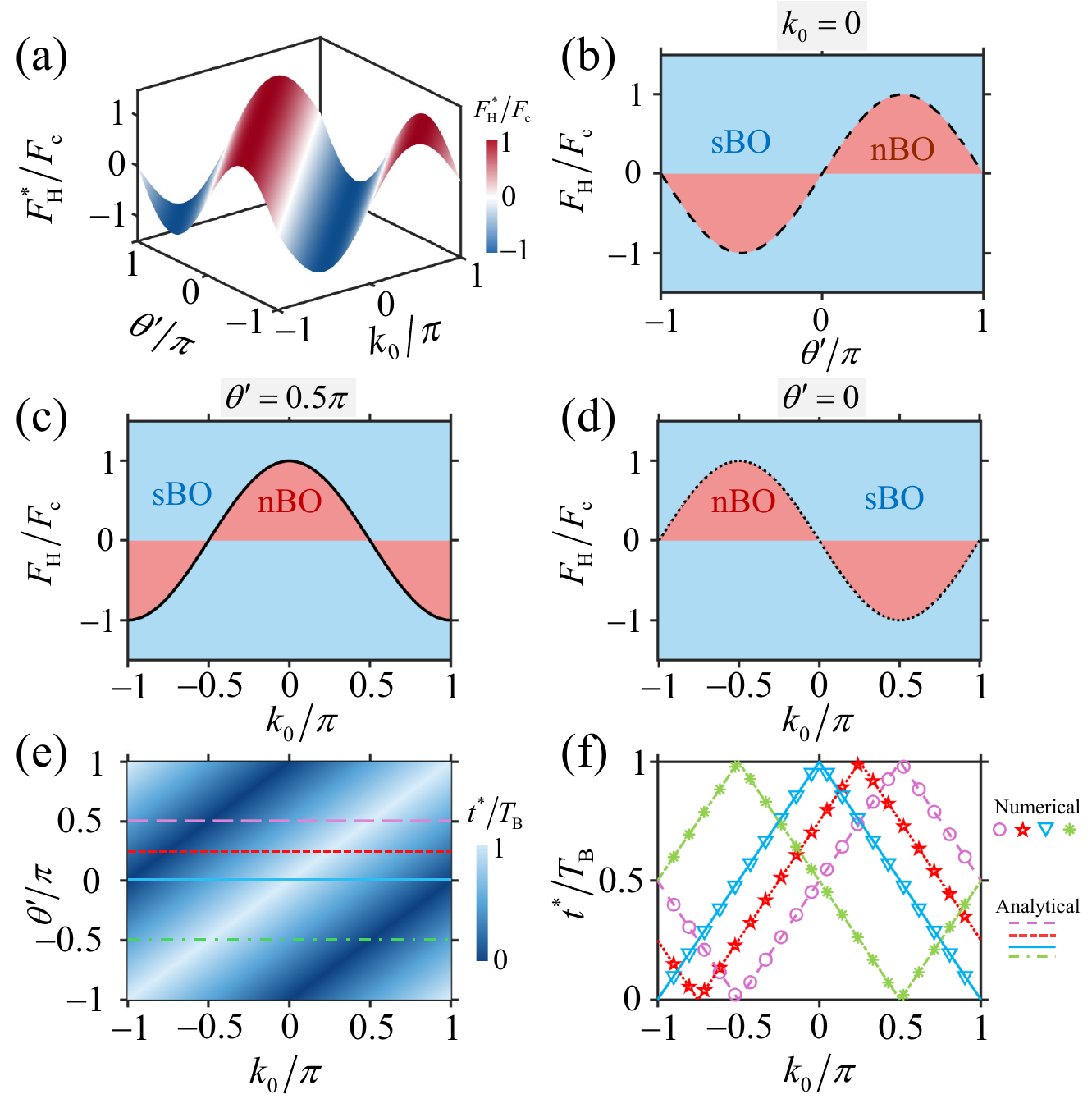}
\caption{(a) Critical force $F_\text{H}^*$ with respect to $k_0$ and $\theta'$. (b) Transition diagram for $k_0=0$. The blue (red) area denotes the numerically founded regime with smooth (nonsmooth) BOs. (c),~(d) Transition diagram for $\theta'=0.5\pi$ (c) and $\theta'=0$ (d). (e) Jump time $t^*$ with respect to $k_0$ and $\theta'$. (f) Predicted jump time $t^*$ with respect $k_0$ for $\theta'=0.5\pi$ (purple dashed line), $\theta'=0.25\pi$ (red dotted line), $\theta'=0$ (blue solid line), and $\theta'=-0.5\pi$ (green dash-dot line). The different symbols denote the numerically founded $t^*$. Other parameters are the same as those in Fig.~\ref{fig1}.} \label{fig3}
\end{figure}

The analytical critical force $F_{\text{H}}^*$ with respect to $\theta'$ and $k_0$ is shown in Fig.~\ref{fig3}(a), which indeed shows a \textit{bifurcation surface} that contains all the bifurcation points. In Fig.~\ref{fig3}(b), we show the transition diagram {as a function of} $\theta'$ while fixing $k_0=0$. The numerically found regions with smooth (nonsmooth) BOs are indicated as blue (red) color, which agree with the predictions in Eq.~(\ref{transition}). The transition boundary is predicted by $F_{\text{H}}^*=F_\text{c}\sin(\theta')$, which is indicated as the dashed line. {In Fig.~\ref{fig3}(c), we then fix $\theta'=0.5\pi$} 
and show the transition diagram {as a function of} $k_0$. The numerically found regions with smooth and nonsmooth BOs are consistent with theoretical predictions in Eq.~(\ref{transition}). The transition between these two regimes agrees well with the predicted value $F_{\text{H}}^*=F_\text{c}\sin(0.5\pi-k_0)=F_\text{c}\cos(k_0)$, as highlighted by the solid line. 
When $|F_\text{H}|<F_\text{c}$, if there exist smooth (nonsmooth) BOs for $F_\text{H}$, it changes to nonsmooth (smooth) BOs for $-F_\text{H}$. While when $|F_\text{H}|>F_\text{c}$, the BOs are always smooth whenever $F_\text{H}$ is positive or negative. We further fix $\theta'=0$, which indicates symmetric non-Hermitian couplings with $J_L=J_R=\mathcal{J}+i\mathcal{J}'$, and plot the diagram in Fig.~\ref{fig3}(d). The correspondence between the numerically obtained smooth and nonsmooth BO regimes match well with the theoretical ones.  

We also examine the jump time $t^*$. We plot the theoretical jump time $t^*$ according to Eq.~(\ref{t_star}) with respect to $k_0$ and $\theta'$ in Fig.~\ref{fig3}(e). {In Fig.~\ref{fig3}(f),} we show the numerically found $t^*$ for different $\theta'$ in different symbols, which agree with the predicted $t^*$ from Eq.~(\ref{t_star}) indicated by lines. In particular, for $\theta'=\pm0.5\pi$ and $k_0=0$, the jump time is $t^*=T_\text{B}/2$, agreeing with the results in Fig.~\ref{fig1}.

Notably, our formulation provides a general framework for NH BOS, even for the case of single-site excitation. We recover the previously found anomalous momentum drift of $F_\text{H}/2$~\cite{graefe2016quasiclassical}, 
{which we can further generalize to }
\begin{align}
    k_{\text{m}}(t)=\theta'-\frac{F_{\text{H}}}{2}t. \label{km}
\end{align}
This equation shows that the maximum momentum is first dominated by the momentum $k=\theta'$ and subsequently drifts at a rate of $-F_{\text{H}}/2$ (see End Matter). Moreover, in a finite NH lattice with unidirectional couplings, we reveal an anomalous wave propagation along the direction with the vanishing hopping strength (see SM \cite{supp}).

\textit{\textcolor{blue}{Experimental outlook}}---The essential ingredients for observing the predicted nonsmooth BOs are the NH couplings, effective constant forces, and finite-width wave packets, all of which are accessible in various platforms. Synthetic dimensions \cite{yu2025comprehensive,yuan2018synthetic,ozawa2019topologicalquantum,lustig2021topological} represent one  promising platform, where NH couplings \cite{weidemann2020topological,wang2021generating,liang2022dynamic,wang2025nonlinear} and effective forces \cite{chen2021real,oliver2023bloch,xie2026quantum} have been engineered independently, enabling observing the time-resolved wave evolution. The predicted nonsmooth-smooth BOs transition can be identified through measuring the wave-packet trajectory, and reversing the force provides a direct test of the nonreciprocal dynamics. Other classical systems including photonic waveguides \cite{2024Observation,sun2024photonic}, acoustic lattices \cite{zhang2021acoustic,gu2022transient,zhong2026observation}, mechanical metamaterials \cite{ghatak2020observation,li2024observation}, and electric circuits \cite{helbig2020generalized} are also possible platforms for demonstrating these predictions. 

\textit{\textcolor{blue}{Conclusion}}---We have established a unified framework for NH BOs in one-dimensional NH lattices and uncovered a dynamical transition between smooth and nonsmooth BOs driven by a momentum-space pitchfork bifurcation. We demonstrate that such a transition can occur in general one-dimensional single-band NH lattices and derive a universal relation for the critical force, which forms a bifurcation surface in parameter space. Remarkably, the bifurcation originates from the extremal structure of the momentum distribution generated by purely linear wave evolution, rather than from nonlinear interactions \cite{jurgensen2021quantized,jurgensen2023quantized,kawabata2025hopf}. Our work therefore reveals a new route to generating nonsmooth transport in linear systems. We anticipate that our framework can be further extended to multiband and higher-dimensional systems, where the effects of interband transitions \cite{breid2006bloch,trompeter2006visual} and quantum geometry \cite{silberstein2020berry,hu2025quantum,montag2026quantum} may give rise to even richer NH wave dynamics.

\begin{acknowledgments}
\textit{Acknowledgments}—--We thank Dr.~Alexander Felski for useful discussions. This work is supported by JSPS KAKENHI Grant No. JP24K00548, JST PRESTO Grant No. JPMJPR2353, JST CREST Grant No. JPMJCR26XA, and Frontier Research in Duo (FRiD) grant from Tohoku University.\\
\end{acknowledgments}


\medskip

\onecolumngrid

\appendix
\medskip


{\centering\textbf{End Matter} \par}
\bigskip
\appendix

\twocolumngrid

\begin{figure}[h!]
\centering
\includegraphics[width=\linewidth]{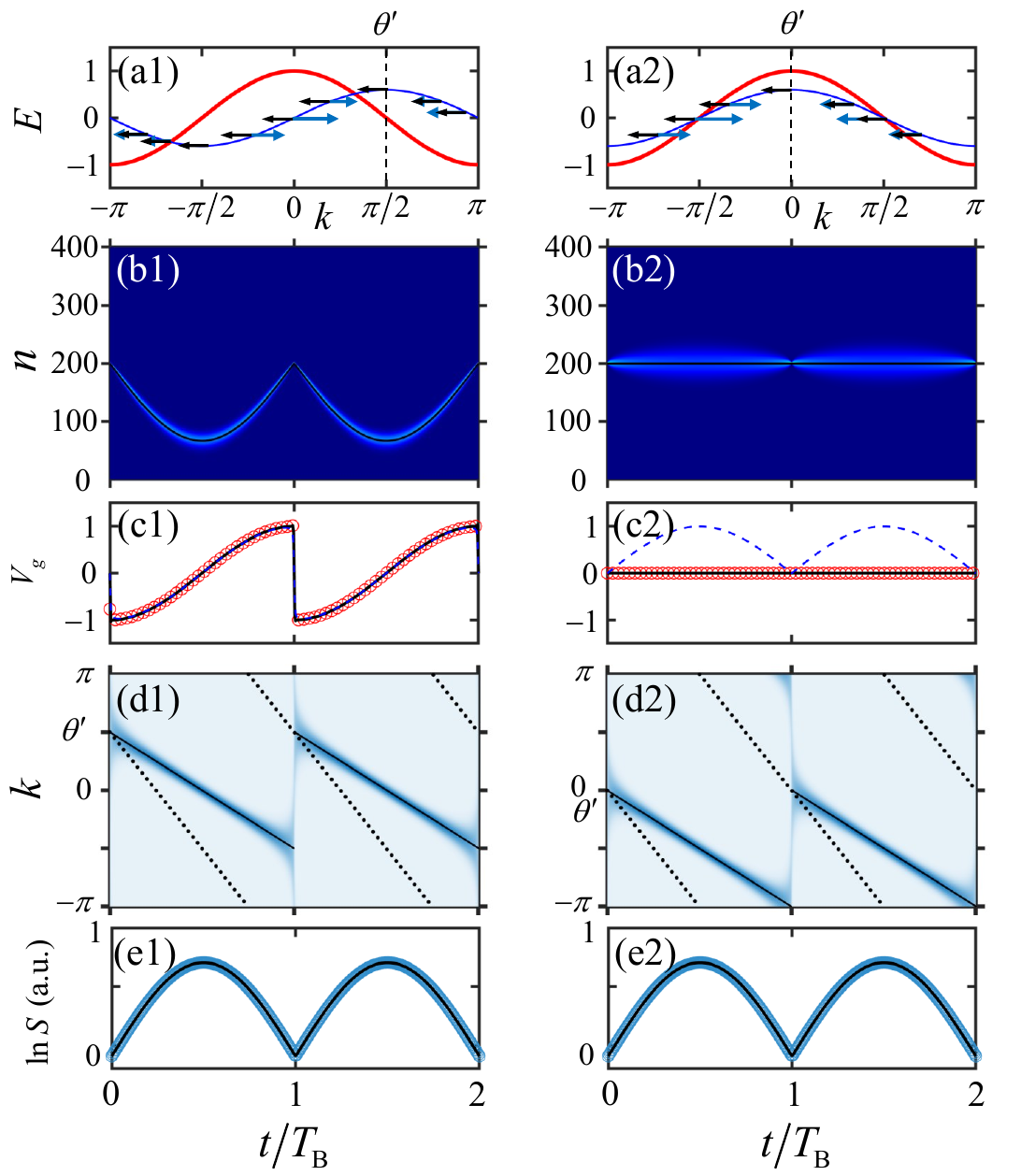}
\caption{(a1-d1) and (a2-d2) Same as Figs.~\ref{fig1} (b1-f1) and~\ref{fig1} (b2-f2), respectively, for the single-site initial state. (e1,e2) Numerical (circles) and theoretical (line) values of the momentum-space amplitude $|S(k_{\text{m}},t)|$. Other parameters are the same as those in Fig.~\ref{fig1}.}
\label{fig4}
\end{figure}
\textit{{Appendix: Single-site excitation}}---We consider the single-site excitation, $\psi_n(0)=\delta_{n,N/2}$, which corresponds to the limiting case $\sigma\rightarrow{0}$ {of a Gaussian wave packet}. In this case, all the momentum components of the band are initially excited and driven by both $F_{\text{H}}$ and $F_{\text{NH}}$ [Figs.~\ref{fig4}(a1) and~\ref{fig4}(a2)]. We fix $F_{\text{H}}=0.015$ and consider two representative cases, $\theta'=\pi/2$ and $\theta'=0$, whose results are shown in Figs.~\ref{fig4}(a1-d1) and~\ref{fig4}(a2-d2), respectively. For $\theta'=\pi/2$, the real-space wave dynamics exhibits periodic oscillations with cusps, with the center of mass oscillating over only left side of the lattice [Fig.~\ref{fig4}(b1)] \cite{peng2022manipulating,graefe2016quasiclassical}, in contrast to the symmetric breathing modes in Hermitian systems \cite{hartmann2004dynamics}. The group velocity exhibits abrupt changes during the time evolution, which is the hallmark of the {nonsmooth BOs} [Fig.~\ref{fig4}(c1)]. For $\theta'=0$, the NH BOs exhibit symmetric breathing modes, with $\langle n(t)\rangle=0$ and $V_g(t)=0$ throughout the evolution (see analytical analysis in SM \cite{supp}). The predicted center of mass $\langle n(t)\rangle$ [Eq.~(\ref{n_average})] and the group velocity $V_g(t)$ [Eq.~(\ref{V})] are in agreement with the numerical results. This anomalous wave-packet dynamics, characterized by a continuously evolving momentum but vanishing group velocity [Figs.~\ref{fig4}(c2) and~\ref{fig4}(d2)], originates from the \textit{anomalous group velocity }$V_A(t)$, which exactly cancels the Hermitian group velocity $\langle dE_R/dk\rangle$, resulting in the vanishing net group velocity [Fig.~\ref{fig4}(c2)]. 

Notably, in both cases, the maximum momentum drifts at a rate of $-F_{\text{H}}/2$  [Figs.~\ref{fig4}(d1) and~\ref{fig4}(d2)], {as previously noted in~\cite{graefe2016quasiclassical}, which is in contrast to Hermitian BO where the momentum drift is at a rate $-F_{\text{H}}$}. The anomalous momentum drift {was previously derived using quasiclassical method in~\cite{graefe2016quasiclassical}, but we now show that this result can also be derived directly from} the dynamics of $S(k,t)$. For the single-site limit $\sigma\rightarrow{0}$, one has $M(k,t)=\Phi_I(k,t)=(4\mathcal{J}'/F_{\text{H}})\sin(F_{\text{H}}t/2)\cos(k+F_{\text{H}}t/2-\theta')$. The maximum momentum $k_{\text{m}}(t)$ is determined by the peak of $|S(k,t)|$, equivalently by the maximum of $\cos(k+F_{\text{H}}t/2-\theta')$, which yields Eq.~(\ref{km}). This equation shows that the maximum momentum is first dominated by the momentum $k=\theta'$ and subsequently drifts at a rate of $-F_{\text{H}}/2$ [Figs.~\ref{fig4}(d1) and~\ref{fig4}(d2)]. This behavior is markedly different from the Hermitian case, where $F_{\text{H}}$ drives the momentum according to $k_{\text{m}}(t)=k_0-F_{\text{H}}t$ [dotted lines in Figs.~\ref{fig4}(d1) and~\ref{fig2}(d2)] \cite{hartmann2004dynamics}. The amplitude $|S(k_{\text{m}},t)|$ first increases and then decreases within each Bloch period, as shown by the blue circles in Figs.~\ref{fig4}(e1) and~\ref{fig2}(e2), matching well with the analytical envelop $(4\mathcal{J}'/F_{\text{H}})\sin(F_{\text{H}}t/2)$ represented by the black solid lines.



\bibliography{reference}
\onecolumngrid

\clearpage

\onecolumngrid

\appendix

\section*{Supplemental Material for: Nonsmooth Bloch Oscillations in Non-Hermitian Systems}

\renewcommand{\thefigure}{S\arabic{figure}}
\setcounter{figure}{0}

\renewcommand{\theequation}{S\arabic{equation}}
\setcounter{equation}{0}














\section{I. Relations between $\mathcal{J,J'},\theta,\theta'$ and $J_{L},J_R$ }
{For one-dimensional single-band tight-binding models with nearest-neighbor hoppings, the band dispersion can be written as}
\begin{align}
E(k)&=E_R(k)+iE_I(k)\nonumber\\
&=2\mathcal{J}\cos(k-\theta)+2i\mathcal{J}'\cos(k-\theta')\nonumber\\
&=(\mathcal{J}e^{-i\theta}+i\mathcal{J}'e^{-i\theta'})e^{ik}+(\mathcal{J}e^{i\theta}+i\mathcal{J}'e^{i\theta'})e^{-ik}. \label{E1}
\end{align}
It can also be written as
\begin{align}
E(k)&=E_R(k)+iE_I(k)\nonumber\\
&=J_Le^{ik}+J_Re^{-ik}\nonumber\\
&=(J_L^{\text{Re}}+iJ_L^{\text{Im}})e^{ik}+(J_R^{\text{Re}}+iJ_R^{\text{Im}})e^{-ik},
\end{align}
where $J_{L}=J_L^{\text{Re}}+iJ_L^{\text{Im}}$ and $J_{R}=J_R^{\text{Re}}+iJ_R^{\text{Im}}$. 

Combining these two equations, we have
\begin{align}
    J_L^{\text{Re}}=\mathcal{J}\cos{\theta}+\mathcal{J}'\sin{\theta'},~J_L^{\text{Im}}=(\mathcal{J}'\cos{\theta'}-\mathcal{J}\sin{\theta}),\nonumber\\
    J_R^{\text{Re}}=\mathcal{J}\cos{\theta}-\mathcal{J}'\sin{\theta'},~J_R^{\text{Im}}=(\mathcal{J}'\cos{\theta'}+\mathcal{J}\sin{\theta}),
\end{align}
and 
\begin{align}
    E_R(k)&=(J_L^{\text{Re}}+J_R^{\text{Re}})\cos{k}+(J_R^{\text{Im}}-J_L^{\text{Im}})\sin{k},\\
    E_I(k)&=(J_L^{\text{Re}}-J_R^{\text{Re}})\sin{k}+(J_L^{\text{Im}}+J_R^{\text{Im}})\cos{k}.
\end{align}

\section{II. Derivation of the extreme momentum}

We consider the wave packet whose real-space profile is $\psi_n (t = 0) = \frac{1}{(2\pi \sigma^2)^{1/4}}e^{-\frac{(n-n_0)^2}{4\sigma^2}}e^{ik_0(n-n_0)}$. In the infinite-size limit, the wavefunction in momentum space is
\begin{align}
    S(k,0)=
    \frac{1}{\sqrt{2\pi}}\int_{-\infty}^\infty \psi_n(0)e^{-ikn}dn
    =
    \left(\frac{2\sigma^2}{\pi}\right)^{1/4}e^{-\sigma^2(k-k_0)^2}e^{-ikn_0}.
\end{align}
The evolution of the wavefunction in momentum space obeys \cite{longhi2015bloch}
\begin{align}
    i\frac{\partial S(k,t)}{\partial t}=E(k)S(k,t)+iF_{\text{H}}\frac{\partial S(k,t)}{\partial k}. \label{partial_S}
\end{align}
The general solution is 
\begin{align}
    S(k,t)=S(k+F_{\text{H}}t,0)e^{-i\int_0^tdt' E(k+F_{\text{H}}t-F_{\text{H}}t')}=S(k+F_{\text{H}}t,0)e^{-i\Phi(k,t)}, \label{S}
\end{align}
where $\Phi(k,t)=\int_0^tdt' E(k+F_{\text{H}}t-F_{\text{H}}t')$. 

Let $q=k+F_{\text{H}}t-F_{\text{H}}t'$, and we have $dq=-F_{\text{H}}dt'$. Therefore,
\begin{align}
    \Phi(k,t)&=\int_0^tdt' E(k+F_{\text{H}}t-F_{\text{H}}t')=-\frac{1}{F_{\text{H}}}\int_{k+F_{\text{H}}t}^{k}dqE(q)=\frac{1}{F_{\text{H}}}\int_{k}^{k+F_{\text{H}}t}dqE(q) \nonumber\\
    &=\frac{1}{F_{\text{H}}}\int_k^{k+F_{\text{H}}t}dqE_R(q)+i\frac{1}{F_{\text{H}}}\int_k^{k+F_{\text{H}}t}dqE_I(q)\nonumber\\
    &= \Phi_R(k,t)+i\Phi_I(k,t),
\end{align}
where 
\begin{align}
    \Phi_{R}(k,t)=\frac{1}{F_{\text{H}}}\int_k^{k+F_{\text{H}}t}dqE_{R}(q)=\frac{1}{F_{\text{H}}}\mathcal{E} _{R}(k+F_{\text{H}}t)-\frac{1}{F_{\text{H}}}\mathcal{E} _{R}(k), \\
    \Phi_{I}(k,t)=\frac{1}{F_{\text{H}}}\int_k^{k+F_{\text{H}}t}dqE_{I}(q)=\frac{1}{F_{\text{H}}}\mathcal{E} _{I}(k+F_{\text{H}}t)-\frac{1}{F_{\text{H}}}\mathcal{E} _{I}(k),
\end{align}
with $\partial\mathcal{E} _{R(I)}(k)/\partial k=E_{R(I)}(k)$. Then we have
\begin{align}
    \frac{\partial \Phi_{R(I)}}{dk}=\frac{E_{R(I)}(k+F_{\text{H}}t)-E_{R(I)}(k)}{F_{\text{H}}}. \label{partial_Phi}
\end{align}
We also have
\begin{align}
    \Phi_{I}(k,t)=\frac{1}{F_{\text{H}}}\mathcal{E} _{I}(k+F_{\text{H}}t)-\frac{1}{F_{\text{H}}}\mathcal{E}_{I}(k)&=\frac{2\mathcal{J}'}{F_{\text{H}}}[\sin(k+F_{\text{H}}t-\theta')-\sin(k-\theta')]\nonumber\\
    &=\frac{4\mathcal{J}'}{F_{\text{H}}}\sin\left(\frac{F_{\text{H}}t}{2}\right)\cos\left(k+\frac{F_{\text{H}}t}{2}-\theta'\right). \label{PHI_I}
\end{align}
The amplitude of the wavefunction in momentum space is 
\begin{align}
    |S(k,t)|=\left(\frac{2\sigma^2}{\pi}\right)^{\frac{1}{4}}e^{-\sigma^2(k+F_{\text{H}}t-k_0)^2}e^{\Phi_I(k,t)}=Ce^{M(k,t)}, \label{Sk}
\end{align}
where $C=\left(\frac{2\sigma^2}{\pi}\right)^{\frac{1}{4}}$ and $M(k,t)=-\sigma^2(k+F_{\text{H}}t-k_0)^2+\Phi_I(k,t)$. The extreme momentum $k_{\text{e}}(t)$ is given by 
\begin{align}
    \frac{\partial M(k,t)}{\partial k}=-2\sigma^2(k+F_{\text{H}}t-k_0)+\frac{\partial \Phi_I(k,t)}{\partial k}=0,\label{partial_X}
\end{align}
which gives 
\begin{align}
\frac{E_I(k+F_{\text{H}}t)-E_I(k)}{ F_{\text{H}}}=2\sigma^2(k+F_{\text{H}}t-k_0),\label{partial_M_SM}
\end{align}
i.e.,
\begin{align}
\frac{E_I(k+F_{\text{H}}t)-E_I(k)}{2\sigma^2 F_{\text{H}}}=k+F_{\text{H}}t-k_0.
\end{align}
This equation determines $k_{\text{e}}(t)$ in a general single-band NH lattice model.

Consider the single-band model with nearest-neighbor hopping with band dispersion in Eq.~(\ref{E1}), we then have
\begin{align}
\frac{\mathcal{J}'}{F_{\text{H}}\sigma^2}[\cos(k+F_{\text{H}}t-\theta')-\cos(k-\theta')]=k+F_{\text{H}}t-k_0,
\end{align}
that is, 
\begin{align}
    \frac{F_\text{c}}{F_{\text{H}}}\sin\left(\frac{F_{\text{H}}t}{2}\right)\sin\left(k+\frac{F_{\text{H}}t}{2}-\theta'\right)=k_0-k-F_{\text{H}}t, \label{ke_SM}
\end{align}
which is Eq.~(8) in the main text. Here $F_\text{c} \equiv \frac{2\mathcal{J}'}{\sigma^2}$.

For the case of $F_{\text{H}}\rightarrow{0}$, the left hand of Eq.~(\ref{partial_M_SM}) is 
\begin{align}
    \frac{1}{F_{\text{H}}}[E_I(k+Ft)-E_I(k)]=t\frac{1}{F_{\text{H}}t}[E_I(k+F_{\text{H}}t)-E_I(k)]\rightarrow{\frac{\Delta E_I}{\Delta k}t},
\end{align}
with $\Delta k=F_{\text{H}}t\rightarrow{0}$. The right hand of Eq.~(\ref{partial_M_SM}) is 
\begin{align}
    2\sigma^2(k+F_{\text{H}}t-k_0)\rightarrow{2\sigma^2(k-k_0)}.
\end{align}
Then Eq.~(\ref{partial_M_SM}) is equivalent to 
\begin{align}
   \frac{\Delta E_I}{\Delta k}t=2\sigma^2(k-k_0),
\end{align}
which gives
\begin{align}
    k_{\text{m}}(t)=k_0+\frac{1}{2\sigma^2} \frac{d E_I}{d k}\bigg|_{k=k_{\text{m}}}t.
\end{align}
This equation denotes the evolution of maximum momentum in NH lattices without forces and has been found previously \cite{he2026anomalous}. It shows that the non-Hermiticity $E_I(k)$ drives the max momentum toward the direction $\frac{d E_I}{d k}$ with increasing gain.

For the Hermitian case with $E_I(k)=0$, Eq.~(\ref{partial_M_SM}) becomes
\begin{align}
    2\sigma^2(k+F_{\text{H}}t-k_0)=0,
\end{align}
which gives the acceleration theorem \cite{hartmann2004dynamics}
\begin{align}
    k_{\text{m}}(t)=k_0-F_{\text{H}}t.
\end{align}

\section{III. Derivation of the center of mass}

In this section, we derive the expression of the center of mass of the wave packet in NH lattices with a  force. The center of mass is expressed as
\begin{align}
    \langle n(t)\rangle=\frac{\sum_n n |\psi_n(t)|^2}{\sum_n  |\psi_n(t)|^2}=\frac{P(t)}{D(t)},
\end{align}
where the numerator and the denominator are defined as $P(t)$ and $D(t)$, respectively. 
 {Assuming that we are in an infinite-size lattice}, the wavefunction in real space can be written as
\begin{align}
    \psi_n(t)=\frac{1}{\sqrt{2\pi}}\int_{-\pi}^\pi S(k,t)e^{ikn}dk.
\end{align}
So, we have
\begin{align}
    |\psi_n(t)|^2=\frac{1}{2\pi}\int_{-\pi}^{\pi}dk\int_{-\pi}^{\pi}dk'S(k,t)S^*(k',t)e^{i(k-k')n}.
\end{align}
Then
\begin{align}
    P(t)&=\sum_n n |\psi_n(t)|^2=\frac{1}{2\pi} \int_{-\pi}^{\pi}dk\int_{-\pi}^{\pi}dk'S(k,t)S^*(k',t)\sum_n ne^{i(k-k')n},\label{P_t} \\
    D(t)&=\sum_n  |\psi_n(t)|^2=\frac{1}{2\pi} \int_{-\pi}^{\pi}dk\int_{-\pi}^{\pi}dk'S(k,t)S^*(k',t)\sum_ne^{i(k-k')n}.\label{D_t}
\end{align}
Since $k\in(-\pi,\pi]$, the Dirac comb function gives
\begin{align}
    \sum_n e^{i(k-k')n}=2\pi\delta(k-k'). \label{Dirac}
\end{align}
Then we have
\begin{align}
 \frac{\partial}{\partial k'}\sum_n e^{i(k-k')n}=-i\sum_nn e^{i(k-k')n}=2\pi \frac{\partial \delta(k-k')}{\partial k'},
\end{align}
which gives
\begin{align}
    \sum_n ne^{i(k-k')n}=i2\pi \frac{\partial \delta(k-k')}{\partial k'}.\label{partial_k_prime}
\end{align}
In the same way, we have
\begin{align}
 \frac{\partial}{\partial k}\sum_n e^{i(k-k')n}=i\sum_nn e^{i(k-k')n}=2\pi \frac{\partial \delta(k-k')}{\partial k},
\end{align}
which gives
\begin{align}
    \sum_n ne^{i(k-k')n}=-i2\pi \frac{\partial \delta(k-k')}{\partial k}. \label{partial_k}
\end{align}
Substituting Eq.~(\ref{partial_k_prime}) into Eq.~(\ref{P_t}), we obtain 
\begin{align}
    P(t)=i\int_{-\pi}^{\pi}dk\int_{-\pi}^{\pi}dk'S(k,t)S^*(k',t)\frac{\partial \delta(k-k')}{\partial k'}.    
\end{align}
Using the property of Dirac function, 
\begin{align}
    \int f(k') \frac{\partial\delta(k-k')}{\partial k'}dk'=-\frac{df(k')}{dk'}\big|_{k'=k},
\end{align}
with $f(k')=S^*(k',t)$. Then we have
\begin{align}
    \int_{-\pi}^{\pi}dk'S^*(k',t)\frac{\partial \delta(k-k')}{\partial k'}=-\frac{\partial S^*(k,t)}{\partial k},  
\end{align}
which gives
\begin{align}
    P(t)=-i\int_{-\pi}^{\pi}dkS(k,t)\frac{\partial S^*(k,t)}{\partial k}.\label{Pt1}    
\end{align}
Similarly, substituting Eq.~(\ref{partial_k}) into Eq.~(\ref{P_t}), we have
\begin{align}
    P(t)=-i\int_{-\pi}^{\pi}dk\int_{-\pi}^{\pi}dk'S(k,t)S^*(k',t)\frac{\partial \delta(k-k')}{\partial k}. 
\end{align}
Using the property of Dirac function,
\begin{align}
    \int f(k) \frac{\partial\delta(k-k')}{\partial k}dk=-\frac{df(k)}{dk}\big|_{k=k'},
\end{align}
$P(t)$ can be rewritten as
\begin{align}
     P(t)=i\int_{-\pi}^{\pi}dk'S^*(k',t)\frac{\partial S(k',t)}{\partial k'}=i\int_{-\pi}^{\pi}dkS^*(k,t)\frac{\partial S(k,t)}{\partial k}. \label{Pt2}
\end{align}
Adding Eq.~(\ref{Pt1}) and Eq.~(\ref{Pt2}) and dividing by tow, we obtain 
\begin{align}
     P(t)=\text{Re} \left[\int_{-\pi}^{\pi}dkS^*(k,t)i\frac{\partial S(k,t)}{\partial k}\right]. 
\end{align}
We can also obtain
\begin{align}
    D(t)=\frac{1}{2\pi} \int_{-\pi}^{\pi}dk\int_{-\pi}^{\pi}dk'S(k,t)S^*(k',t)\sum_ne^{i(k-k')n}=\int_{-\pi}^{\pi}dk |S(k,t)|^2.
\end{align}
Then center of mass is then 
\begin{align}
    \langle n(t)\rangle=\frac{P(t)}{D(t)}=\frac{\text{Re} \left[\int_{-\pi}^{\pi}dkS^*(k,t)i\frac{\partial S(k,t)}{\partial k}\right]}{\int_{-\pi}^{\pi}dk |S(k,t)|^2}.
\end{align}
We first calculate $\frac{\partial S(k,t)}{\partial k }$. From Eq.~(\ref{S}), we obtain 
\begin{align}
    \frac{\partial S}{\partial k}=\frac{\partial S_0}{\partial k}e^{-i\Phi}-i\frac{\partial\Phi}{\partial k}S_0e^{-i\Phi},
\end{align}
where $S_0=S(k+F_{\text{H}}t,0)$. Substituting Eq.~(\ref{partial_Phi}) into this equation, we obtain 
\begin{align}
    \frac{\partial S}{\partial k}=\frac{\partial S_0}{\partial k}e^{-i\Phi_R}e^{\Phi_I}-i\left[\frac{E_{R}(k+F_{\text{H}}t)-E_{R}(k)}{F_{\text{H}}} +i\frac{E_{I}(k+F_{\text{H}}t)-E_{I}(k)}{F_{\text{H}}}\right]S_0e^{-i\Phi_R}e^{\Phi_I}.
\end{align}
Therefore, 
\begin{align}
    S^*(k,t)i\frac{\partial S(k,t)}{\partial k}&=S_0^*e^{i\Phi_R}e^{\Phi_I}\left\{ i\frac{\partial S_0}{\partial k}e^{-i\Phi_R}e^{\Phi_I}+\left[\frac{E_{R}(k+F_{\text{H}}t)-E_{R}(k)}{F} +i\frac{E_{I}(k+F_{\text{H}}t)-E_{I}(k)}{F_{\text{H}}}\right]S_0e^{-i\Phi_R}e^{\Phi_I}\right\}\nonumber\\
    &=iS_0^*\frac{\partial S_0}{\partial k}e^{2\Phi_I}+|S_0|^2e^{2\Phi_I}\left[\frac{E_{R}(k+F_{\text{H}}t)-E_{R}(k)}{F} +i\frac{E_{I}(k+F_{\text{H}}t)-E_{I}(k)}{F_{\text{H}}}\right]\nonumber\\
&=iS_0^*\frac{\partial S_0}{\partial k}e^{2\Phi_I}+|S_0|^2e^{2\Phi_I}\frac{E_{R}(k+F_{\text{H}}t)-E_{R}(k)}{F_{\text{H}}} +i|S_0|^2e^{2\Phi_I}\frac{E_{I}(k+F_{\text{H}}t)-E_{I}(k)}{F_{\text{H}}}.  
\end{align}
On the other hand,
\begin{align}
    S_0=S(k+F_{\text{H}}t,0)=Ce^{-\sigma^2(k+F_{\text{H}}t-k_0)^2}e^{-i(k+F_{\text{H}}t)n_0}.
\end{align}
Then,
\begin{align}
    \frac{\partial S_0}{\partial k}=-2C\sigma^2(k+F_{\text{H}}t-k_0)e^{-\sigma^2(k+F_{\text{H}}t-k_0)^2}e^{-i(k+F_{\text{H}}t)n_0}+Ce^{-\sigma^2(k+F_{\text{H}}t-k_0)^2}[-in_0e^{-i(k+F_{\text{H}}t)n_0}].
\end{align}
So we have
\begin{align}
    iS_0^*\frac{\partial S_0}{\partial k}&=iCe^{-\sigma^2(k+F_{\text{H}}t-k_0)^2}e^{i(k+F_{\text{H}}t)n_0}\frac{\partial S_0}{\partial k}\nonumber\\
    &=-2iC^2\sigma^2(k+F_{\text{H}}t-k_0)e^{-2\sigma^2(k+F_{\text{H}}t-k_0)^2}+C^2n_0e^{-2\sigma^2(k+F_{\text{H}}t-k_0)^2}.
\end{align}
Then,
\begin{align}
    P(t)&=\text{Re} \left[\int_{-\pi}^{\pi}dkS^*(k,t)i\frac{\partial S(k,t)}{\partial k}\right]=\int_{-\pi}^{\pi}\text{Re} \left[iS_0^*\frac{\partial S_0}{\partial k}e^{2\Phi_I}\right]+\int_{-\pi}^{\pi}
|S_0|^2e^{2\Phi_I}\frac{E_{R}(k+F_{\text{H}}t)-E_{R}(k)}{F_{\text{H}}}\nonumber\\
&=\int_{-\pi}^{\pi}dkC^2n_0e^{-2\sigma^2(k+F_{\text{H}}t-k_0)^2}e^{2\Phi_I}+\int_{-\pi}^{\pi}dk|S_0|^2e^{2\Phi_I}\frac{E_{R}(k+F_{\text{H}}t)-E_{R}(k)}{F_{\text{H}}}\label{P_real}
\end{align}
Therefore, 
\begin{align}
    \langle n(t)\rangle
    &=\frac{\int_{-\pi}^{\pi}dkC^2n_0e^{-2\sigma^2(k+F_{\text{H}}t-k_0)^2}e^{2\Phi_I}+\int_{-\pi}^{\pi}dk|S_0|^2e^{2\Phi_I}\frac{E_{R}(k+F_{\text{H}}t)-E_{R}(k)}{F_{\text{H}}}}{\int_{-\pi}^{\pi}dk|S_0|^2e^{2\Phi_I}}\nonumber\\
    &=\frac{\int_{-\pi}^{\pi}dkC^2n_0e^{-2\sigma^2(k+F_{\text{H}}t-k_0)^2}}{\int_{-\pi}^{\pi}dkC^2e^{-2\sigma^2(k+F_{\text{H}}t-k_0)^2}}+\frac{\int_{-\pi}^{\pi}dk|S_0|^2e^{2\Phi_I}\frac{E_{R}(k+F_{\text{H}}t)-E_{R}(k)}{F_{\text{H}}}}{\int_{-\pi}^{\pi}dk|S_0|^2e^{2\Phi_I}}\nonumber\\
    &=n_0+\left\langle \frac{E_{R}(k+F_{\text{H}}t)-E_{R}(k)}{F_{\text{H}}}\right\rangle, \label{n}
\end{align}
where the momentum-space average for a function $f(k)$ is defined by 
\begin{align}
    \langle f(k)\rangle=\frac{\int_{-\pi}^{\pi}dk|S_0|^2e^{2\Phi_I}f(k)}{\int_{-\pi}^{\pi}dk|S_0|^2e^{2\Phi_I}}=\frac{\int_{-\pi}^{\pi}dk|S(k,t)|^2f(k)}{\int_{-\pi}^{\pi}dk|S(k,t)|^2}.
\end{align}
Eq.~(\ref{n}) is Eq.~(5) of the main text.

\section{IV. Derivation of the group velocity}
The group velocity of a wave packet is defined by
\begin{align}
V_g(t)=\frac{d\langle n(t)\rangle}{dt}=\frac{\dot{P}D-P\dot{D}}{D^2},
\end{align}
where 
\begin{align}
     P(t)&=\text{Re} \left[\int_{-\pi}^{\pi}dkS^*(k,t)i\frac{\partial S(k,t)}{\partial k}\right]=\int_{-\pi}^{\pi}dk\text{Re}\left[S^*(k,t)i\frac{\partial S(k,t)}{\partial k}\right]=\int_{-\pi}^{\pi}dk\text{Re}\mathcal{P},\nonumber\\
     D(t)&=\int_{-\pi}^{\pi}dk |S(k,t)|^2,
\end{align}
where have defined $\mathcal{P}=S^*(k,t)i\frac{\partial S(k,t)}{\partial k}$.

Then we have
\begin{align}
    \dot{\mathcal{P}}=i\frac{\partial S^*}{\partial t}\frac{\partial S}{\partial k}+iS^* \frac{\partial}{\partial t} \frac{\partial S}{\partial k}. \label{P_dot}
\end{align}
From Eq.~(\ref{partial_S}), we have
\begin{align}
    \frac{\partial S}{\partial t}=-iES+F_{\text{H}}\frac{\partial S}{\partial k}, \label{S_partial}
\end{align}
which gives
\begin{align}
    \frac{\partial S^*}{\partial t}=iE^*S^*+F_{\text{H}}\frac{\partial S^*}{\partial k}. \label{S_star_partial}
\end{align}
So the first part of Eq.~(\ref{P_dot}) is 
\begin{align}
    i\frac{\partial S^*}{\partial t}\frac{\partial S}{\partial k}=i\frac{\partial S}{\partial k}\left(iE^*S^*+F_{\text{H}}\frac{\partial S^*}{\partial k} \right)=-E^*S^*\frac{\partial S}{\partial k}+iF_{\text{H}}\frac{\partial S}{\partial k}\frac{\partial S^*}{\partial k}.\label{iS_first}
\end{align}
On then other hand, the second term of Eq.~(\ref{P_dot}) is 
\begin{align}
    iS^*\frac{\partial}{\partial t} \frac{\partial S}{\partial k}&=iS^*\frac{\partial}{\partial k} \frac{\partial S}{\partial t}=iS^*\frac{\partial}{\partial k}\left(-iES+F_{\text{H}}\frac{\partial S}{\partial k}\right)\nonumber\\
    &=S^*\frac{d E}{d k}S+S^*E\frac{\partial S}{\partial k}+iS^*F_{\text{H}}\frac{\partial}{\partial k}\frac{\partial S}{\partial k}. \label{iS_second}
\end{align}
Combining Eq.~(\ref{iS_first}) and Eq.~(\ref{iS_second}), we have
\begin{align}
    \dot{\mathcal{P}}=i\frac{\partial S^*}{\partial t}\frac{\partial S}{\partial k}+iS^* \frac{\partial}{\partial t} \frac{\partial S}{\partial k}&=-E^*S^*\frac{\partial S}{\partial k}+iF_{\text{H}}\frac{\partial S}{\partial k}\frac{\partial S^*}{\partial k}+S^*\frac{d E}{d k}S+S^*E\frac{\partial S}{\partial k}+iS^*F_{\text{H}}\frac{\partial}{\partial k}\frac{\partial S}{\partial k}\nonumber\\
    &=2iE_IS^*\frac{\partial S}{\partial k}+|S|^2\frac{d E}{d k}+iF_{\text{H}}\frac{\partial S}{\partial k}\frac{\partial S^*}{\partial k}+iS^*F_{\text{H}}\frac{\partial }{\partial k}\frac{\partial S}{\partial k}.
\end{align}
So,
\begin{align}
     \dot{P}(t)=\text{Re}\left(\int_{-\pi}^{\pi} dk\dot{\mathcal{P}}\right)=\text{Re}\left[\int_{-\pi}^{\pi}dk\left(2iE_IS^*\frac{\partial S}{\partial k}+|S|^2\frac{d E}{d k}+iF_{\text{H}}\frac{\partial S}{\partial k}\frac{\partial S^*}{\partial k}+iS^*F_{\text{H}}\frac{\partial }{\partial k}\frac{\partial S}{\partial k}\right)\right].
\end{align}
We find that
\begin{align}
    \int_{-\pi}^{\pi}dk\left(iF_{\text{H}}\frac{\partial S}{\partial k}\frac{\partial S^*}{\partial k}+iS^*F_{\text{H}}\frac{\partial }{\partial k}\frac{\partial S}{\partial k}\right)=iF_{\text{H}}\int_{-\pi}^{\pi}dk\frac{\partial}{\partial k} \left(S^*\frac{\partial S}{\partial k}\right)
    =iF_{\text{H}}\left(S^*\frac{\partial S}{\partial k}\right)\bigg|_{-\pi}^{\pi}
    =0,
\end{align}
where we have used the property of $S(k+\pi,t)=S(k-\pi,t)$.

Therefore, we have
\begin{align}
     \dot{P}(t)=\text{Re}\left(\int_{-\pi}^{\pi} dk\dot{\mathcal{P}}\right)=\int_{-\pi}^{\pi}dk2E_I\text{Re}\left(iS^*\frac{\partial S}{\partial k}\right)+\int_{-\pi}^{\pi} dk|S|^2\frac{d E_R}{d k}.
\end{align}
On the other hand, using Eq.~(\ref{S_partial}) and Eq.~(\ref{S_star_partial}), we have
\begin{align}
     \dot{D}(t)&=\int_{-\pi}^{\pi}dk \left(\frac{\partial S}{\partial t}S^*+\frac{\partial S^*}{\partial t}S\right)=\int_{-\pi}^{\pi}dk \left(-iES+F_{\text{H}}\frac{\partial S}{\partial k}\right)S^*+\int_{-\pi}^{\pi}dk \left(iE^*S^*+F_{\text{H}}\frac{\partial S^*}{\partial k}\right)S\nonumber\\
     &=\int_{-\pi}^{\pi}dk \left(2E_I|S|^2+F_{\text{H}}\frac{\partial S}{\partial k}S^*+F_{\text{H}}\frac{\partial S^*}{\partial k}S\right)\nonumber\\
     &=2\int_{-\pi}^{\pi}dkE_I|S|^2,
\end{align}
where we have used 
\begin{align}
    \int_{-\pi}^{\pi}dk \left(F_{\text{H}}\frac{\partial S}{\partial k}S^*+F_{\text{H}}\frac{\partial S^*}{\partial k}S\right)=F_{\text{H}}\int_{-\pi}^{\pi}dk \frac{\partial}{\partial k}\left(SS^*\right)=F_{\text{H}}|S|^2\bigg|_{-\pi}^{\pi}=0.
\end{align}
Then,
\begin{align}
    V_g(t)=\frac{\dot{P}D-P\dot{D}}{D^2}&=\frac{\left[\int_{-\pi}^{\pi}dk2E_I\text{Re}\left(iS^*\frac{\partial S}{\partial k}\right)+\int_{-\pi}^{\pi} dk|S|^2\frac{d E_R}{d k} \right]\int_{-\pi}^{\pi}dk |S|^2}{\left(\int_{-\pi}^{\pi}dk |S|^2\right)^2}\nonumber\\
    &-\frac{\int_{-\pi}^{\pi}dk\text{Re} (S^*i\frac{\partial S}{\partial k})2\int_{-\pi}^{\pi}dkE_I|S|^2}{\left(\int_{-\pi}^{\pi}dk |S|^2\right)^2}\nonumber\\
    &=\frac{\int_{-\pi}^{\pi}dk2E_I\text{Re}\left(iS^*\frac{\partial S}{\partial k}\right)}{\int_{-\pi}^{\pi}dk |S|^2}+\frac{\int_{-\pi}^{\pi} dk|S|^2\frac{d E_R}{d k}}{\int_{-\pi}^{\pi}dk |S|^2}-\frac{\int_{-\pi}^{\pi}dk\text{Re} (S^*i\frac{\partial S}{\partial k})2\int_{-\pi}^{\pi}dkE_I|S|^2}{(\int_{-\pi}^{\pi}dk |S|^2)(\int_{-\pi}^{\pi}dk |S|^2)}. \label{v}
\end{align}
Based on Eq.~(\ref{P_real}), the first term is
\begin{align}
    \frac{\int_{-\pi}^{\pi}dk2E_I\text{Re}\left(iS^*\frac{\partial S}{\partial k}\right)}{\int_{-\pi}^{\pi}dk |S|^2}&=2\frac{\int_{-\pi}^{\pi}dkE_I\left[C^2n_0e^{-2\sigma^2(k+F_{\text{H}}t-k_0)^2}e^{2\Phi_I}+|S_0|^2e^{2\Phi_I}\frac{E_{R}(k+F_{\text{H}}t)-E_{R}(k)}{F_{\text{H}}}\right]}{\int_{-\pi}^{\pi}dk |S|^2}\nonumber\\
    &=2n_0\langle E_I\rangle+2\left\langle E_I \frac{E_{R}(k+F_{\text{H}}t)-E_{R}(k)}{F_{\text{H}}}\right\rangle. \label{v1}
\end{align}
The second term is 
\begin{align}
    \frac{\int_{-\pi}^{\pi} dk|S|^2\frac{d E_R}{d k}}{\int_{-\pi}^{\pi}dk |S|^2}=\left\langle \frac{d E_R}{d k}\right\rangle.\label{v2}
\end{align}
The last term is 
\begin{align}
    -\frac{\int_{-\pi}^{\pi}dk\text{Re} (S^*i\frac{\partial S}{\partial k})2\int_{-\pi}^{\pi}dkE_I|S|^2}{(\int_{-\pi}^{\pi}dk |S|^2)(\int_{-\pi}^{\pi}dk |S|^2)}
    &=-\frac{\int_{-\pi}^{\pi}dk\text{Re} (S^*i\frac{\partial S}{\partial k})}{\int_{-\pi}^{\pi}dk |S|^2}\frac{2\int_{-\pi}^{\pi}dkE_I|S|^2}{\int_{-\pi}^{\pi}dk |S|^2}\nonumber\\
    &=-\frac{\int_{-\pi}^{\pi}dk\left[C^2n_0e^{-2\sigma^2(k+F_{\text{H}}t-k_0)^2}e^{2\Phi_I}+|S_0|^2e^{2\Phi_I}\frac{E_{R}(k+F_{\text{H}}t)-E_{R}(k)}{F_{\text{H}}}\right]}{\int_{-\pi}^{\pi}dk |S|^2}\frac{2\int_{-\pi}^{\pi}dkE_I|S|^2}{\int_{-\pi}^{\pi}dk |S|^2}\nonumber\\
    &=-2n_0\langle E_I\rangle-2\left\langle \frac{E_{R}(k+F_{\text{H}}t)-E_{R}(k)}{F_{\text{H}}}\right\rangle \langle E_I \rangle.\label{v3}
\end{align}
Substituting Eqs.~(\ref{v1})-(\ref{v3}) into Eq.~(\ref{v}), the group velocity is then 
\begin{align}
      V_g(t)=\left\langle \frac{d E_R}{d k}\right\rangle+2\left[\left\langle \frac{E_{R}(k+F_{\text{H}}t)-E_{R}(k)}{F_{\text{H}}}E_I\right\rangle -\left\langle \frac{E_{R}(k+F_{\text{H}}t)-E_{R}(k)}{F_{\text{H}}}\right\rangle \langle E_I \rangle\right], \label{V_SM}
\end{align}
which is Eq.~(6) in the main text.


\section{V. Finite-size effect with PBCs}

In this section, we analyze the finite-size effects for the wave-packet dynamics in lattices with PBC. For a finite lattice with the size $N$ and $n_0=N/2$, the momentum-space wavefunction is approximately equal to \cite{he2026anomalous}
\begin{align}
    \psi_k(0)=\frac{1}{\sqrt{2\pi}}\sum_n\psi_n(0)e^{-ikn}
    &\approx\frac{1}{\sqrt{2\pi}}\int_{1}^N \psi_n(0)e^{-ikn}dn\nonumber\\
    &\approx S(k,0)
    \frac{1}{2}
    \left\{\text{erf}\left( \frac{N}{4\sigma} + i\sigma (k-k_0)\right)
    -
    \text{erf}\left( -\frac{N}{4\sigma}+i\sigma (k-k_0)\right)
    \right\}\nonumber\\
    &=S(k,0)\Theta(k),\label{S_theta}
\end{align}
where
\begin{align}
    \Theta(k)=\frac{1}{2}
    \left[\text{erf}\left( a + bi\right)
    +
    \text{erf}\left( a-bi\right)
    \right]=\text{Re}[\text{erf}(a+ib)],
\end{align} 
with $a \equiv \frac{N}{4\sigma}$ and $b \equiv \sigma (k-k_0)$.

\begin{figure}[h!]
\centering
\includegraphics[width=0.8\linewidth]{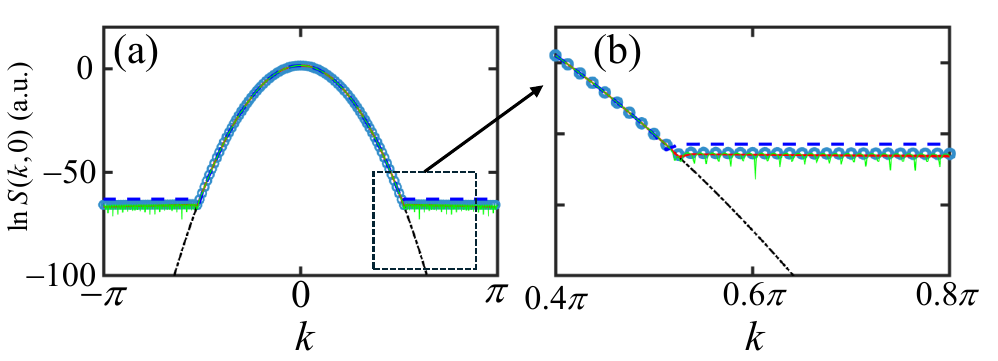}
\caption{(a) Momentum-space wavefunction. The blue circles represent numerical $\psi_k(0)$ for the finite discrete lattice. The black dash-dotted line denotes the counterpart $S(k,0)$ for the infinite case. The green solid line denotes $S(k,0)\Theta(k)$ for the finite continuous case based on Eq.~(\ref{S_theta}), and the red solid line denotes $S(k,0)\Theta(k)$ for the finite discrete lattice based on Eq.~(\ref{S_theta}). The blue dashed line denote the approximate result $S(k,0)\Theta(k)$ with $\Theta(k)$ used in Eq.~(\ref{Theta}). (b) Enlarged area highlighted by the box in (a).}
\label{Fig_PBC_finite}
\end{figure}

For a large $z$, the error function $\text{erf}(z)$ can be written as
\begin{align}
    \text{erf}(z)=\frac{2}{\sqrt{\pi}}\int_0^z e^{-u^2}du=1-\frac{2}{\sqrt{\pi}}\int_z^{\infty} e^{-u^2}du\approx 1-\frac{2}{\sqrt{\pi}}\frac{e^{-z^2}}{2z}= 1-\frac{e^{-z^2}}{\sqrt{\pi}z}.
\end{align}
Since $|a + bi|$ is large, we then have the approximation
\begin{align}
    \Theta(k)=\text{Re}[\text{erf}(a+ib)]\approx1-\frac{e^{b^2-a^2}}{\sqrt{\pi}}\frac{a\cos(2ab)-b\sin(2ab)}{a^2+b^2}=1-\frac{e^{b^2-a^2}}{\sqrt{\pi}}\frac{\cos(2ab+\arctan(\frac{b}{a}))}{\sqrt{a^2+b^2}}=1-\frac{e^{b^2-a^2}}{\sqrt{\pi}}\mathcal{A}.
\end{align}
Here $\mathcal{A}=\frac{\cos(2ab+\arctan(\frac{b}{a}))}{\sqrt{a^2+b^2}}$ contributes to the oscillations, and does not affect the shape of the function $\Theta(k)$ [see Fig.~\ref{Fig_PBC_finite}]. We hence we make a reasonable approximation that 
\begin{align}
    \Theta(k)\approx1-\frac{e^{b^2-a^2}}{\sqrt{\pi}}=1-\frac{e^{\sigma^2(k-k_0)^2-\frac{N^2}{16\sigma^2}}}{\sqrt{\pi}}. \label{Theta}
\end{align}

In Fig.~\ref{Fig_PBC_finite}, we compare the numerically obtained and approximate momentum-space wavefunctions. We see that the approximate function $S(k,0)\Theta(k)$ (blue dashed line) nearly matches the numerical result, $\psi_k(0)=\frac{1}{\sqrt{2\pi}}\sum_n\psi_n(0)e^{-ikn}$ (blue circles), indicating that Eq.~(\ref{Theta}) provides a reasonable approximation.

Based on the above analysis, we rewrite the modified wavefunction in momentum space as $S(k,t)=S(k+F_{\text{H}}t,0)\Theta(k+F_{\text{H}}t)e^{-i\Phi(k,t)}=S'(k,t)e^{i\varphi}$, where $S'(k,t)=Ce^{-\sigma^2(k+F_{\text{H}}t-k_0)^2}\Theta(k+F_{\text{H}}t)e^{\Phi_I(k,t)}$ and $\varphi=-\Phi_R(k,t)-(k+F_{\text{H}}t)n_0$ denote the amplitude and phase factor, respectively. The amplitude can be written as 
\begin{align}
    S'(k,t)&=Ce^{-\sigma^2(k+F_{\text{H}}t-k_0)^2}\Theta(k+F_{\text{H}}t)e^{\Phi_I}\nonumber\\
    &=Ce^{-\sigma^2(k+F_{\text{H}}t-k_0)^2}\left[1-\frac{e^{\sigma^2(k+F_{\text{H}}t-k_0)^2-\frac{N^2}{16\sigma^2}}}{\sqrt{\pi}}\right]e^{\Phi_I}\nonumber\\
    &=C\left[e^{-\sigma^2(k+F_{\text{H}}t-k_0)^2}-\frac{e^{-\frac{N^2}{16\sigma^2}}}{\sqrt{\pi}}\right]e^{\Phi_I}.
\end{align}
Then the extreme values $k_{\text{e}}(t)$ is determined by ${\partial S'(k,t)}/{\partial k}=0$, which gives
\begin{align}
    -2\sigma^2(k+F_{\text{H}}t-k_0)e^{-\sigma^2(k+F_{\text{H}}t-k_0)^2}e^{\Phi_I}+e^{-\sigma^2(k+F_{\text{H}}t-k_0)^2} \frac{\partial\Phi_I}{\partial k}e^{\Phi_I}-\frac{1}{\sqrt{\pi}}e^{-\frac{N^2}{16\sigma^2}}\frac{\partial\Phi_I}{\partial k}e^{\Phi_I}=0,
\end{align}
that is,
\begin{align}
    \frac{\Theta(k+F_{\text{H}}t)}{2\sigma^2}\frac{\partial\Phi_I}{\partial k}=k+F_{\text{H}}t-k_0. \label{ke_PBC}
\end{align}

{For PBC, we implement the effect of a force via time-dependent vector potential as}
\begin{align}
        \tilde{H}=\sum_n J_Le^{iA(t)}|n\rangle\langle n+1|+J_Re^{-iA(t)}|n+1\rangle\langle n|,
\end{align}
where $A(t)=-F_{\text{H}}t$. Although the lattice is finite, the momentum-space wavefunction still approximately obeys Eq.~(\ref{S_partial}), and the momentum can be treated as a continuous variable, provided that the size $N$ is not too small.

\begin{figure}[ht!]
\centering
\includegraphics[width=0.7 \linewidth]{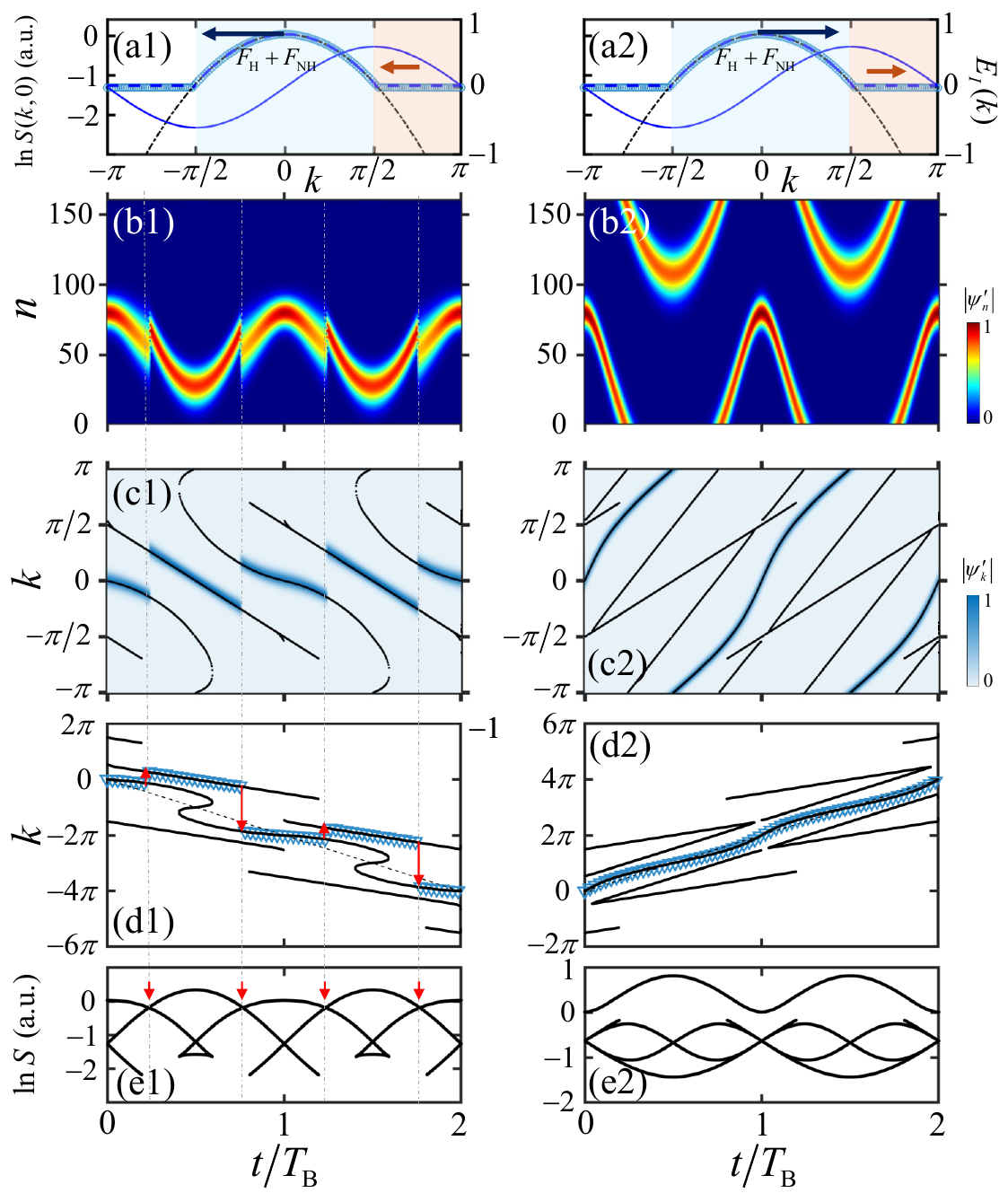}
\caption{(a1,a2) Initial momentum-space wavefunction. Blue solid circles (blue dashed lines) represent the numerical (analytical approximation) results, while the black dash-dotted lines denote the counterpart in the infinite limit. The solid blue lines denote $E_I(k)$. (b1,b2)-(e1,e2) Same as those in Fig.~1 of the main text. }
\label{fig3_SP}
\end{figure}

In Figs.~\ref{fig3_SP}(a1) and~\ref{fig3_SP}(a2), we see that initial wavefunction $\text{ln}|S(k,0)|$ has flat tails near the Brillouin-zone boundaries (blue dashed lines), in contrast to the quadratic profile exhibited by the infinite counterpart (black dash-dotted line). Compared to the infinite case, the finite-size effects introduce additional solutions of $k_{\text{e}}(t)$, giving rise to more complex wave-packet jumps. For $F_{\text{H}}>0$, the initial momentum drifts toward $-k$ direction while undergoing loss, as indicated by the blue arrow in Fig.~\ref{fig3_SP}(a1). After a finite time, the momentum jumps to another solution originating from the flat tail of the momentum distribution, which undergoes gain [red arrow in Fig.~\ref{fig3_SP}(a1)] and subsequently drifts at a rate $-F_{\text{H}}/2$ [Figs.~\ref{fig3_SP}(c1) and~\ref{fig3_SP}(d1)]. As time further increases, the dominant momentum jumps back, due to the decrease (increase) in the amplitude of the current (initial) wave-packet momentum branch [Fig.~\ref{fig3_SP}(e1)]. The momentum-space jump gives rise to nonsmooth and discontinuous real-space dynamics, characterized by periodic \textit{NH wave-packet jumps} in Fig.~\ref{fig3_SP}(b1) \cite{he2026anomalous}. In contrast, for $F_{\text{H}}<0$, the initial wavefunction in momentum space drifts toward $+k$ direction while experiencing gain [Fig.~\ref{fig3_SP}(a2)], and therefore remains dominant throughout the evolution [Figs.~\ref{fig3_SP}(c2,d2,e2)]. As a result, the real-space dynamics becomes smooth [Fig.~\ref{fig3_SP}(b2)]. 

\begin{figure}[h!]
\centering
\includegraphics[width=0.8 \linewidth]{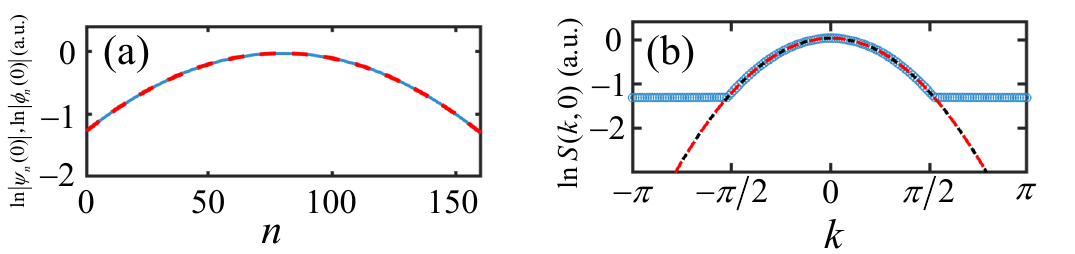}
\caption{(a) Initial wave packet in real space. Blue solid line denotes $\psi_n(0)$, while red dashed line denotes $\phi_n(0)$. (b) The momentum-space wavefunctions of $\psi_n(0)$ (blue circles) and $\phi_n(0)$ (red dashed line). The black dash-dotted line denotes the counterpart in the infinite-size limit.}
\label{Fig_PBC_gaussian}
\end{figure}

We further analyze another type of initial Gaussian wave packet with the form
\begin{align}
    \phi_n(0)=\frac{1}{(2\pi\sigma^2)^{1/4}}\sum_me^{-\frac{(n-n_0+mN)^2}{4\sigma^2}}e^{ik_0(n-n_0+mN)},
\end{align}
which is different from the truncated Gaussian wave packet $\psi_n(0)$. In Fig.~\ref{Fig_PBC_gaussian}, we plot the real-space and the corresponding  momentum-space wavefunctions. We see that $\phi_n(0)$ {agrees very well} with $\psi_n(0)$. However, in momentum space, these two wave packets exhibit difference near the edge of the Brillouian zone. Different from the momentum-space wavefunction of $\psi_n(0)$, which shows a flat tail near the edge of Brillouian zone (blue circles), the momentum-space wavefunction of $\phi_n(0)$ is still Gaussian and {agrees well} with the counterpart of infinite case with the quadratic function (red dashed line). {Because of} the disappearance of the flat tail in momentum space, the jump from the momentum near $\pi/2$ no longer occurs. 

\begin{figure}[ht!]
\centering
\includegraphics[width=0.8 \linewidth]{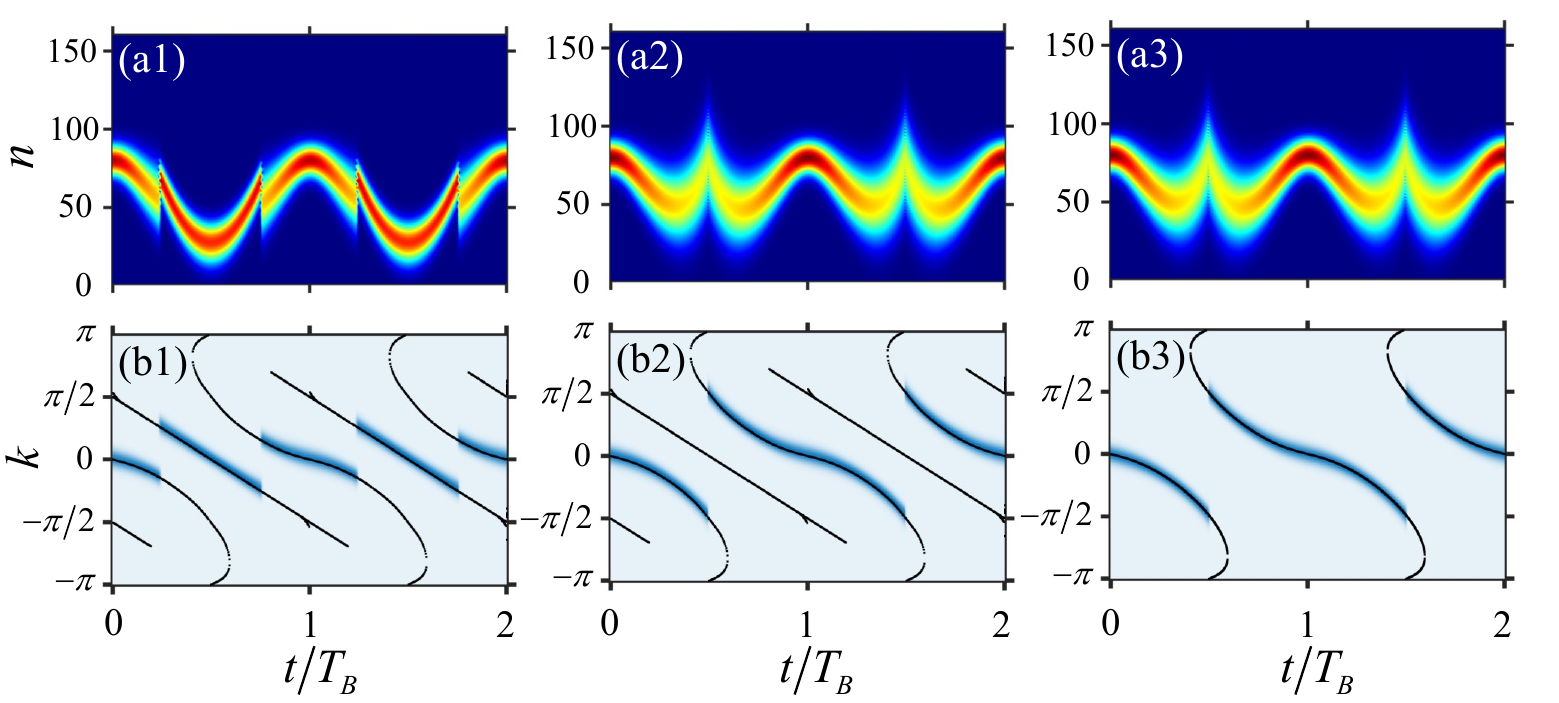}
\caption{(a1-a3) Real-space wave-packet dynamics. (b1-b3) Momentum-space dynamics, where the solid lines
denote $k_{\text{e}}(t)$. (a1,b1) With initial Gaussian wave packet $\psi_n(0)$, where  $k_{\text{e}}(t)$ is obtained based Eq.~(\ref{ke_PBC}). (a2,b2) With initial Gaussian wave packet $\phi_n(0)$, where  $k_{\text{e}}(t)$ is obtained based Eq.~(\ref{ke_PBC}). (a3,b3) With initial Gaussian wave packet $\psi_n(0)$, but for infinite case, where the dynamics is obtained from the evolution of $S(k,t)$ using Eq.~(\ref{partial_S}).  $k_{\text{e}}(t)$ is obtained based Eq.~(\ref{ke_SM}). $F_{\text{H}}=0.015$, and other parameters are the sames as those used in Fig.~\ref{fig3_SP}.}
\label{Fig_PBC_compare}
\end{figure}

In Fig.~\ref{Fig_PBC_compare}, we compare the wave dynamics for different cases. We see that the wave dynamics for the initial wave packet with $\phi_n(0)$ differ from those for $\psi_n(0)$, but are the same as those in the infinite case. The wave-packet jumps induced by the boundary effects disappear for $\phi_n(0)$, and hence the corresponding real-space dynamics coincide with those in the infinite case.

\clearpage
\section{VI. Transition between the nonsmooth and smooth BOs}
In this section, we derive the critical force $F_{\text{H}}^*$ and time $t^*$ for the transition between the smooth and nonsmooth BOs. Based on Eqs. (\ref{PHI_I}-\ref{Sk}), we obtain
\begin{align}
    M(k,t)=-\sigma^2(k+F_{\text{H}}t-k_0)^2+\frac{4\mathcal{J}'}{F_{\text{H}}} \sin\left(\frac{F_{\text{H}}t}{2}\right)\cos\left(k+\frac{F_{\text{H}}t}{2}-\theta'\right),
\end{align}
which can be written as
\begin{align}
    \frac{M(k,t)}{\sigma^2}=-(X-G)^2+2A(t)\cos\left(X\right), \label{M_X}
\end{align}
with
\begin{align}
    X&= k+\frac{F_{\text{H}}}{2}t-\theta ', \nonumber\\
    G&=k_0-\frac{F_{\text{H}}}{2}t-\theta',\nonumber\\
    A(t)&=\frac{F_\text{c}}{F_{\text{H}}}\sin\left(\frac{F_{\text{H}}t}{2}\right).
\end{align}

Let $Y=X-G$, Eq.~(\ref{M_X}) becomes
\begin{align}
    \frac{M(k,t)}{\sigma^2}=-Y^2+2A(t)\cos\left(Y+G\right).\label{M_Y}
\end{align}

We now analyze when the function $M(k,t)$ can have multiple extreme values. From Fig.~2 of the main text, when $M(k,t)$ starts to have two peaks, it is symmetric about one central value which we assume $G^*$. Then, $G^*$ should satisfy the following condition
\begin{align}
    \cos(G^*+Y)=\cos(G^*-Y),
\end{align}
i.e.,
\begin{align}
    \cos(G^*+Y)-\cos(G^*-Y)=-2\sin(G^*)\sin(Y),
\end{align}
which gives
\begin{align}
    \sin(G^*)=0.
\end{align}
Therefore, 
\begin{align}
    G^*=n\pi, n=0,\pm1,\pm2...
\end{align}

(i) When $G^*=2n\pi$,
\begin{align}
    \frac{M(k,t)}{\sigma^2}=-Y^2+2A(t)\cos\left(Y+G^*\right)=-Y^2+2A(t)\cos\left(Y\right).
\end{align}
Then, 
\begin{align}
    \frac{1}{\sigma^2}\frac{\partial M}{\partial Y}=-2Y-2A(t)\sin(Y),\nonumber\\
    \frac{1}{\sigma^2}\frac{\partial^2 M}{\partial Y^2}=-2-2A(t)\cos(Y).
\end{align}
At $Y=0$, 
\begin{align}
  \frac{1}{\sigma^2}\frac{\partial M}{\partial Y}=0,~\frac{1}{\sigma^2} \frac{\partial^2 M}{\partial Y^2}\bigg|_{Y=0}=-2(1+A)<0,
\end{align}
which means that $Y=0$ is always the local maximum, and it cannot be spitted to two peaks. 

(ii) When $G^*=(2n+1)\pi$,
\begin{align}
    \frac{M(k,t)}{\sigma^2}=-Y^2+2A(t)\cos\left(Y+G^*\right)=-Y^2-2A(t)\cos\left(Y\right).
\end{align}
Then, 
\begin{align}
    \frac{1}{\sigma^2}\frac{\partial M}{\partial Y}=-2Y+2A(t)\sin(Y),\nonumber\\
    \frac{1}{\sigma^2}\frac{\partial^2 M}{\partial Y^2}=-2+2A(t)\cos(Y).
\end{align}
At $Y=0$, 
\begin{align}
    \frac{1}{\sigma^2}\frac{\partial M}{\partial Y}\bigg|_{Y=0}=0,~
    \frac{1}{\sigma^2}\frac{\partial^2 M}{\partial Y^2}\bigg|_{Y=0}=2(A-1).
\end{align}
This shows the existence of transition at $A=1$. When $A<1$, $Y=0$ is the local maximum; while when $A>1$, $Y=0$ becomes local minimum, meanwhile, two peaks appear. Therefore, the transition requires $G^*=(2n+1)\pi$.

In this case, $\partial M/\partial Y=0$ gives the following equation that determines $k_{\text{e}}$,
\begin{align}
    Y=A\sin(Y).
\end{align}
When $A<1$, there is only one solution with $Y=0$. When $A>1$, there are three solutions $Y=0,~\pm Y_0$, which are determined by
\begin{align}
    \frac{1}{A}=\frac{\sin (Y_0)}{Y_0}.
\end{align}

We rewrite $G=k_0-\frac{F_{\text{H}}}{2}t-\theta'=-\Delta-\frac{F_{\text{H}}}{2}t$, where $\Delta=\theta'-k_0\in[-\pi,\pi]$.

For $F_{\text{H}}>0$, $G$ decreases with $t$. The transition occurs when $G$ decreases to $-\pi$. Since $-\frac{F_{\text{H}}}{2}t\in[-\pi,0]$ during one BO period, it requires $\Delta\in[0,\pi]$. Therefore the transition condition is 
\begin{align}
   G= -\Delta-\frac{F_{\text{H}}}{2}t^*=-\pi,
\end{align}
which gives
\begin{align}
    t^*=\left(1-\frac{\Delta}{\pi}\right)T_\text{B}.\label{t_star_g}
\end{align}
Then 
\begin{align}
A(t^*)=\frac{F_\text{c}}{F_{\text{H}}}\sin\left(\frac{F_{\text{H}}t^*}{2}\right)=\frac{F_\text{c}}{F_{\text{H}}}\sin\left(\pi-\Delta\right)=\frac{F_\text{c}}{F_{\text{H}}}\sin(\theta'-k_0).
\end{align}
The nonsmooth BOs occur when $A(t^*)>1$, i.e.,
\begin{align}
    F_{\text{H}}<F_\text{c}\sin(\theta'-k_0). \label{F_g}
\end{align}

For $F_{\text{H}}<0$, $G$ increases with $t$. The transition occurs when $G$ increases to $\pi$. Since $-\frac{F_{\text{H}}}{2}t\in[0,\pi]$ during one BO period, it requires $\Delta\in[-\pi,0]$. Therefore the transition condition is 
\begin{align}
   G= -\Delta-\frac{F_{\text{H}}}{2}t^*=\pi,
\end{align}
which gives
\begin{align}
    t^*=\frac{T_\text{B}}{\pi}(\Delta+\pi)=\left(1+\frac{\Delta}{\pi}\right)T_\text{B}.\label{t_star_l}
\end{align}

Then 
\begin{align}
A(t^*)=\frac{F_\text{c}}{F_{\text{H}}}\sin\left(\frac{F_{\text{H}}t^*}{2}\right)=\frac{F_\text{c}}{F_{\text{H}}}\sin\left(\pi-\Delta\right)=\frac{F_\text{c}}{F_{\text{H}}}\sin(\theta'-k_0).
\end{align}
The nonsmooth BOs occur when $A(t^*)>1$, i.e.,
\begin{align}
    F_{\text{H}}>F_\text{c}\sin(\theta'-k_0). \label{F_l}
\end{align}

Based on Eqs.~(\ref{F_g}) and (\ref{F_l}), the general critical force for nonsmooth BOs is 
\begin{align}
    F_{\text{H}}^* =F_\text{c}\sin(\theta'-k_0).
\end{align}
The regime for nonsmooth BOs is 
\begin{align}
    F_{\text{H}}<F_{\text{H}}^*,~~ 
 \text{for}~~~F_{\text{H}}>0, \nonumber\\  
F_{\text{H}}>F_{\text{H}}^*,~~ 
 \text{for}~~~F_{\text{H}}<0.
\end{align}

Based on Eqs.~(\ref{t_star_g}) and (\ref{t_star_l}), the transition time for any $F_{\text{H}}$ is 
\begin{align}
    t^*=\left(1-\frac{|\Delta|}{\pi}\right)T_\text{B}.
\end{align}

Since $Y=k+F_{\text{H}}t-k_0$, the solutions obtained by $Y=0,\pm Y_0$, i.e.,
\begin{align}
    k_{\text{e}}(t^*)=k_0-F_{\text{H}}t^*+Y=\{k_0-F_{\text{H}}t^*,~k_0-F_{\text{H}}t^*+Y_0,~k_0-F_{\text{H}}t^*-Y_0\},
\end{align}
where $Y_0$ is determined by
\begin{align}
    \frac{1}{A(t^*)}=\frac{F_{\text{H}}\sigma^2}{2\mathcal{J}'\sin(\theta'-k_0)}=\frac{\sin (Y_0)}{Y_0}.
\end{align}

For $\theta'=\pi/2$ and $k_0=0$, we have
\begin{align}
    F_{\text{H}}^*=F_\text{c}\sin(\theta'-k_0)=F_\text{c}=\frac{2\mathcal{J}'}{\sigma^2},\nonumber\\
    t^*=\left(1-\frac{|\theta'-k_0|}{\pi}\right)T_\text{B}=\frac{T_\text{B}}{2}. \label{F_star}
\end{align}


We then investigate the influence of the initial Gaussian wave packet width $\sigma$, and show that the transition from nonsmooth to smooth BOs can be also observed. Based on Eq.~(\ref{F_star}), the critical value for $\sigma$ is 
\begin{align}
    \sigma_\text{c}=\sqrt{\frac{2\mathcal{J'}}{|F_{\text{H}}|}.
    }
\end{align}

\begin{figure}[h!]
\centering
\includegraphics[width=\linewidth]{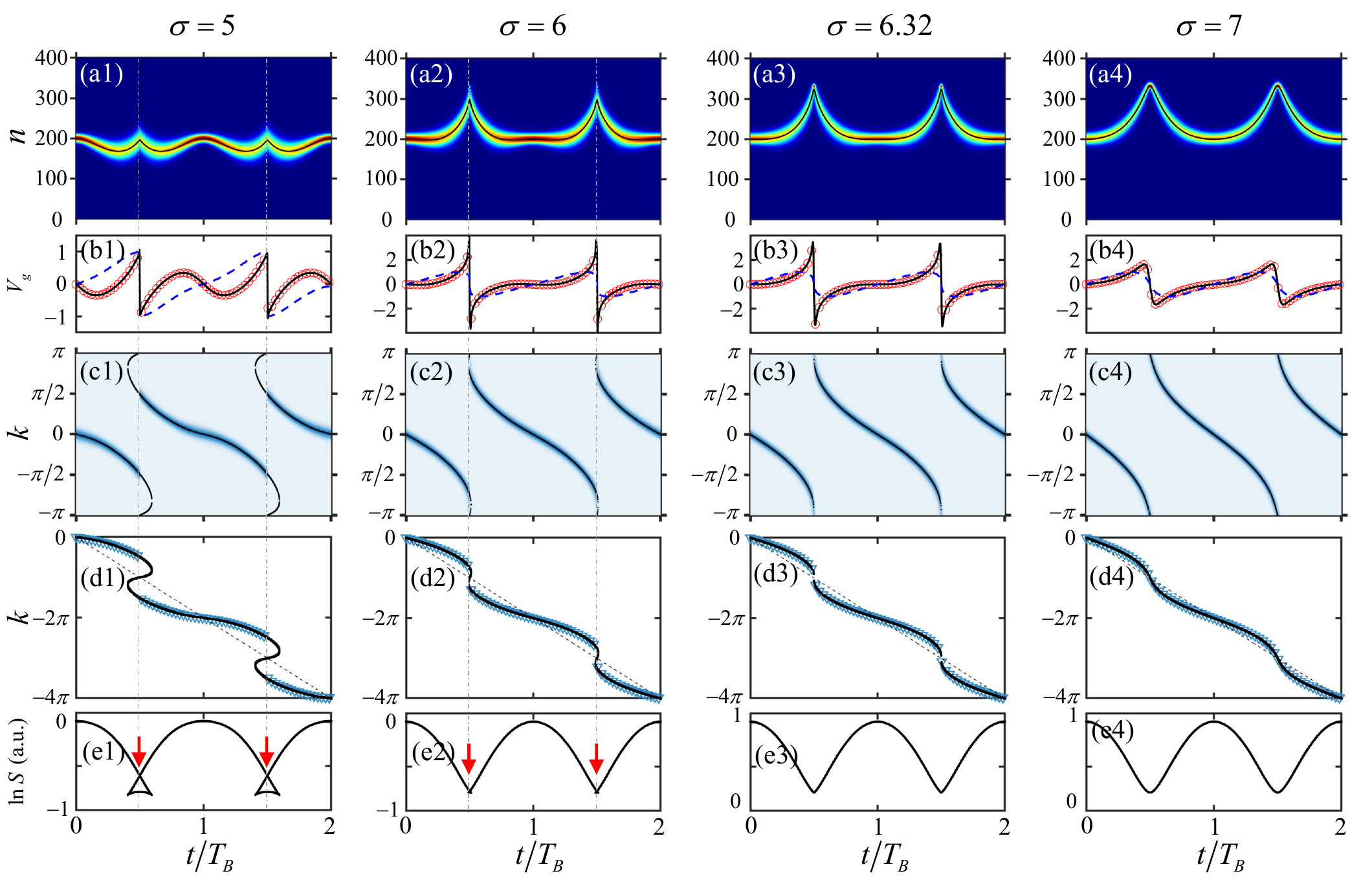}
\caption{(a1-a4) Real-space wave-packet dynamics, where the black solid lines denote the predicted center of mass trajectory $\langle n(t)\rangle$ obtained from Eq.~(\ref{n}). (b1-b4) Numerical (red circles) and the predicted group velocity $V_g(t)$ from Eq.~(\ref{V_SM}) (black solid lines), and the Hermitian counterpart $\langle dE_R/dk\rangle$ (blue dashed lines). (c1-c4) Momentum-space dynamics, where the solid lines denote $k_{\text{e}}(t)$. (d1-d4) Same as (c1-c4), but plotted in the multiple Brillouin zone. The blue triangles represent the numerically extracted $k_{\text{m}}(t)$. (e1-e4) Amplitude of the momentum-space wavefunction evaluated at $k_{\text{e}}(t)$, i.e., $|S(k_{\text{e}},t)|$. Parameters are $\mathcal{J}=0.5,~\mathcal{J}'=0.3,~\theta=0,~\theta_k'=\pi/2,~k_0=0,N=400$,~and $n_0=N/2$. $F_{\text{H}}=0.015$.}
\label{Fig_infinite_F_sigma_p}
\end{figure}

\begin{figure}[h!]
\centering
\includegraphics[width=\linewidth]{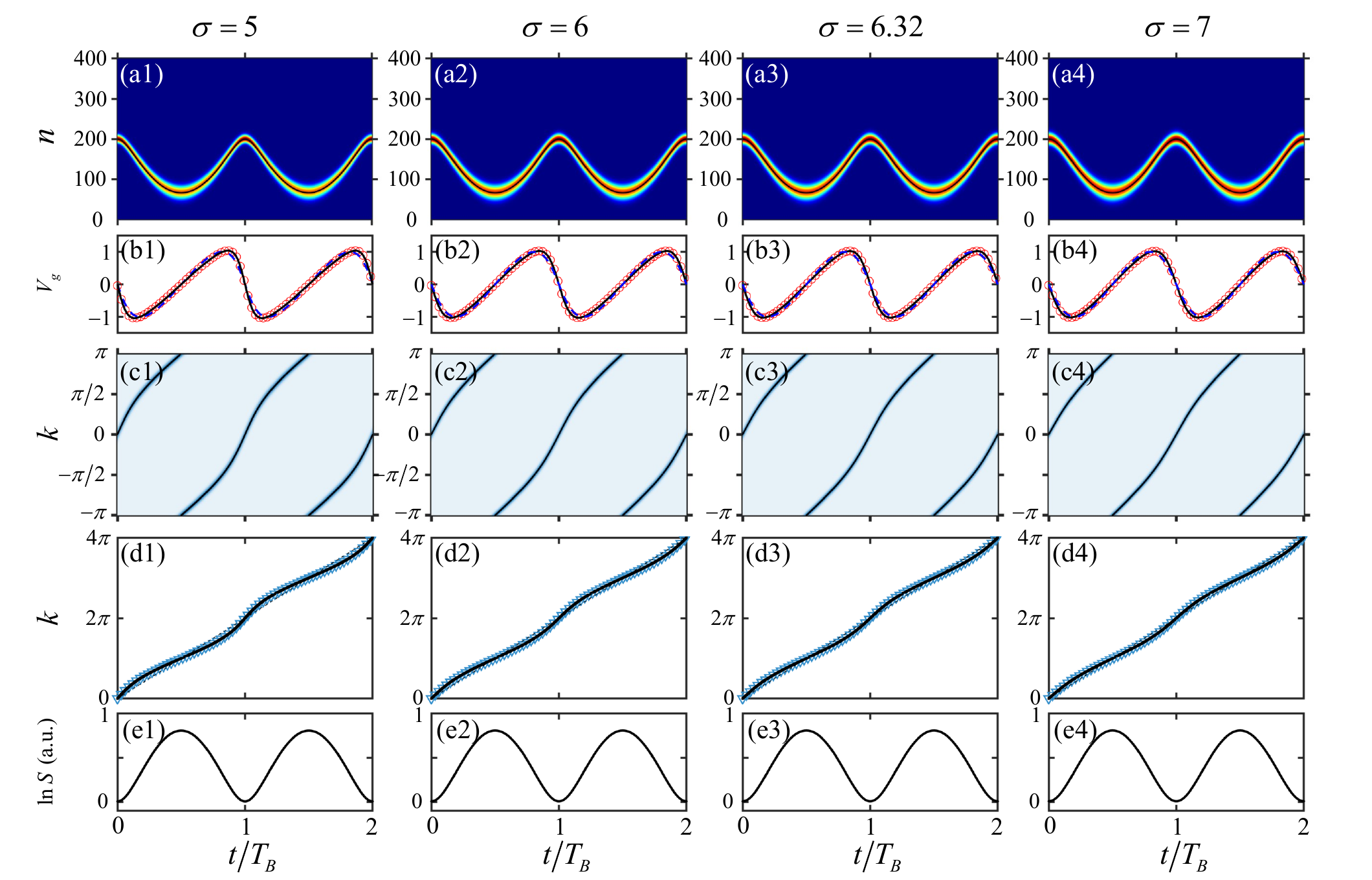}
\caption{Same as those in Fig.~\ref{Fig_infinite_F_sigma_p} with $F_{\text{H}}=-0.015$.}
\label{Fig_infinite_F_sigma_n}
\end{figure}

For the parameters used in the main text, $F_{\text{H}}=0.015,\mathcal{J'}=0.3$, the critical width is $\sigma_\text{c}=\sqrt{\frac{2\mathcal{J'}}{|F_{\text{H}}|}
    }\approx6.32$. Figures~\ref{Fig_infinite_F_sigma_p} and \ref{Fig_infinite_F_sigma_n} plot the results with different $\sigma$ for $F_\text{H}>0$ and $F_\text{H}<0$, respectively. We observe that the BOs are always smooth for $F_\text{H}<0$, whereas for $F_\text{H}>0$, they exhibit a transition from nonsmooth to smooth behavior with increasing $\sigma$. This transition occurs at $\sigma=\sigma_c\approx6.32$ where $k_{\text{e}}(t)$ changes from multiple solutions to a single solution [see Fig.~\ref{Fig_infinite_F_sigma_p} (d3)]. Consequently, the momentum evolution becomes continuous [Fig.~\ref{Fig_infinite_F_sigma_p} (c3)], leading to smooth wave dynamics in real space [Figs.~\ref{Fig_infinite_F_sigma_p} (a3,b3)].  

\clearpage

\begin{figure}[h!]
\centering
\includegraphics[width=0.6\linewidth]{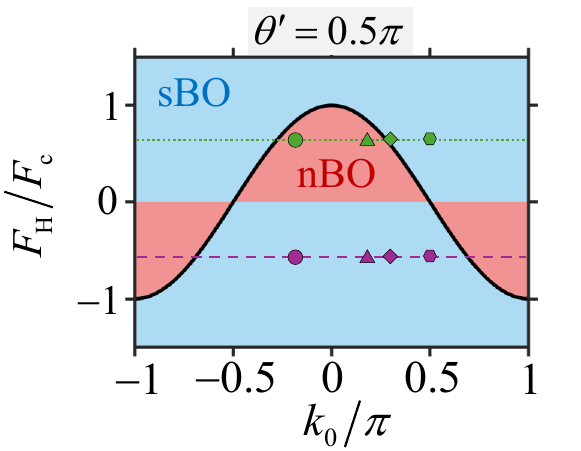}
\caption{Same as Fig.~3(c) in the main text. The green dotted line denote $F_\text{H}=0.015=0.625F_\text{c}$, where the green circle, triangle, diamond, and hexagon denote $k_0=-0.2\pi$, $k_0=0.2\pi$, $k_0=0.3\pi$, and $k_0=0.5\pi$, respectively. The purple dashed line denote $F_\text{H}=-0.015=-0.625F_\text{c}$, where the purple circle, triangle, diamond, and hexagon denote $k_0=-0.2\pi$, $k_0=0.2\pi$, $k_0=0.3\pi$, and $k_0=0.5\pi$, respectively.}
\label{Fig_infinite_k0}
\end{figure}

\begin{figure}[h!]
\centering
\includegraphics[width=\linewidth]{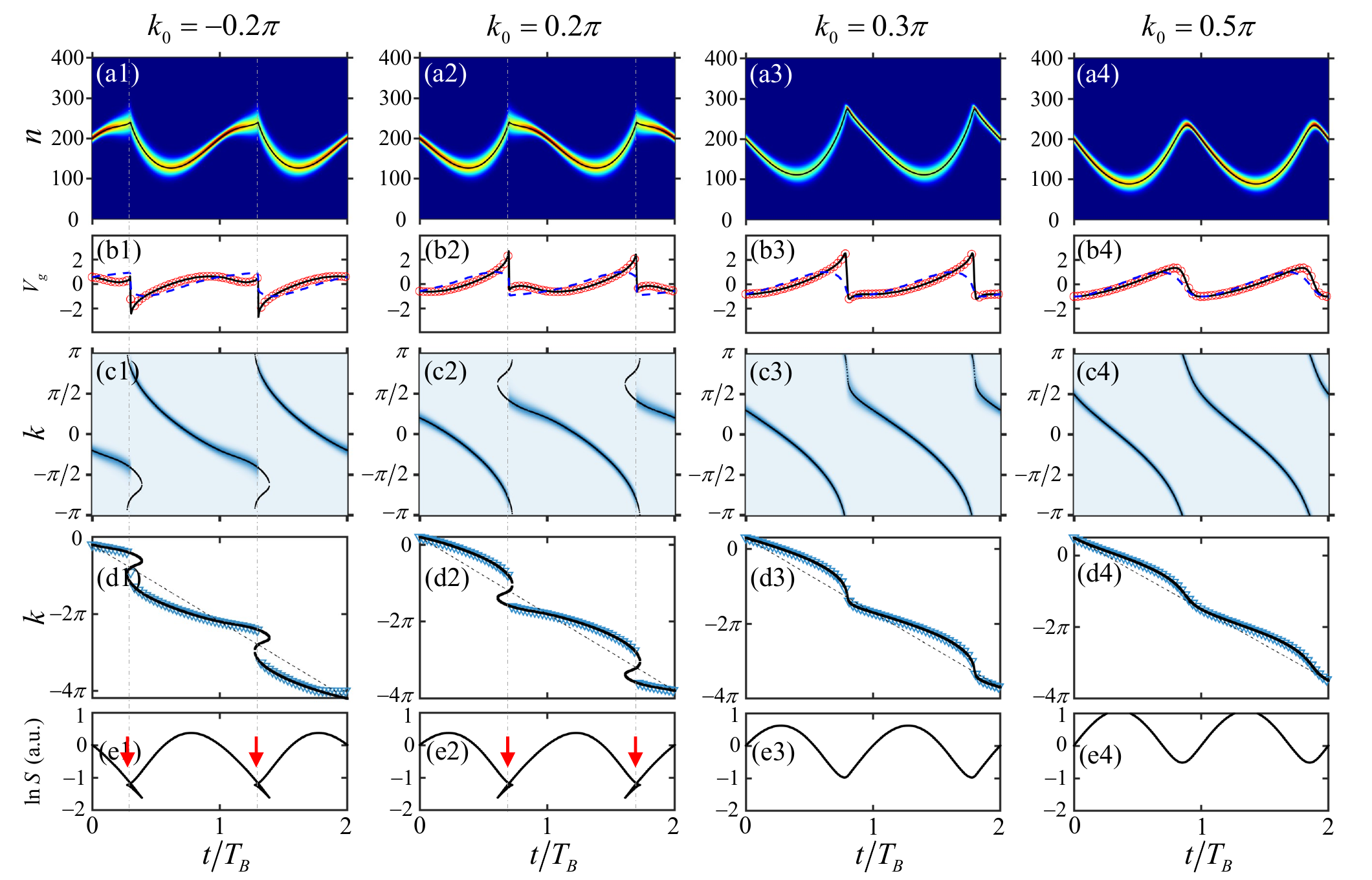}
\caption{Same as those in Fig.~\ref{Fig_infinite_F_sigma_p} with $F_{\text{H}}=0.015$.}
\label{Fig_infinite_F_k0_p}
\end{figure}

\begin{figure}[h!]
\centering
\includegraphics[width=\linewidth]{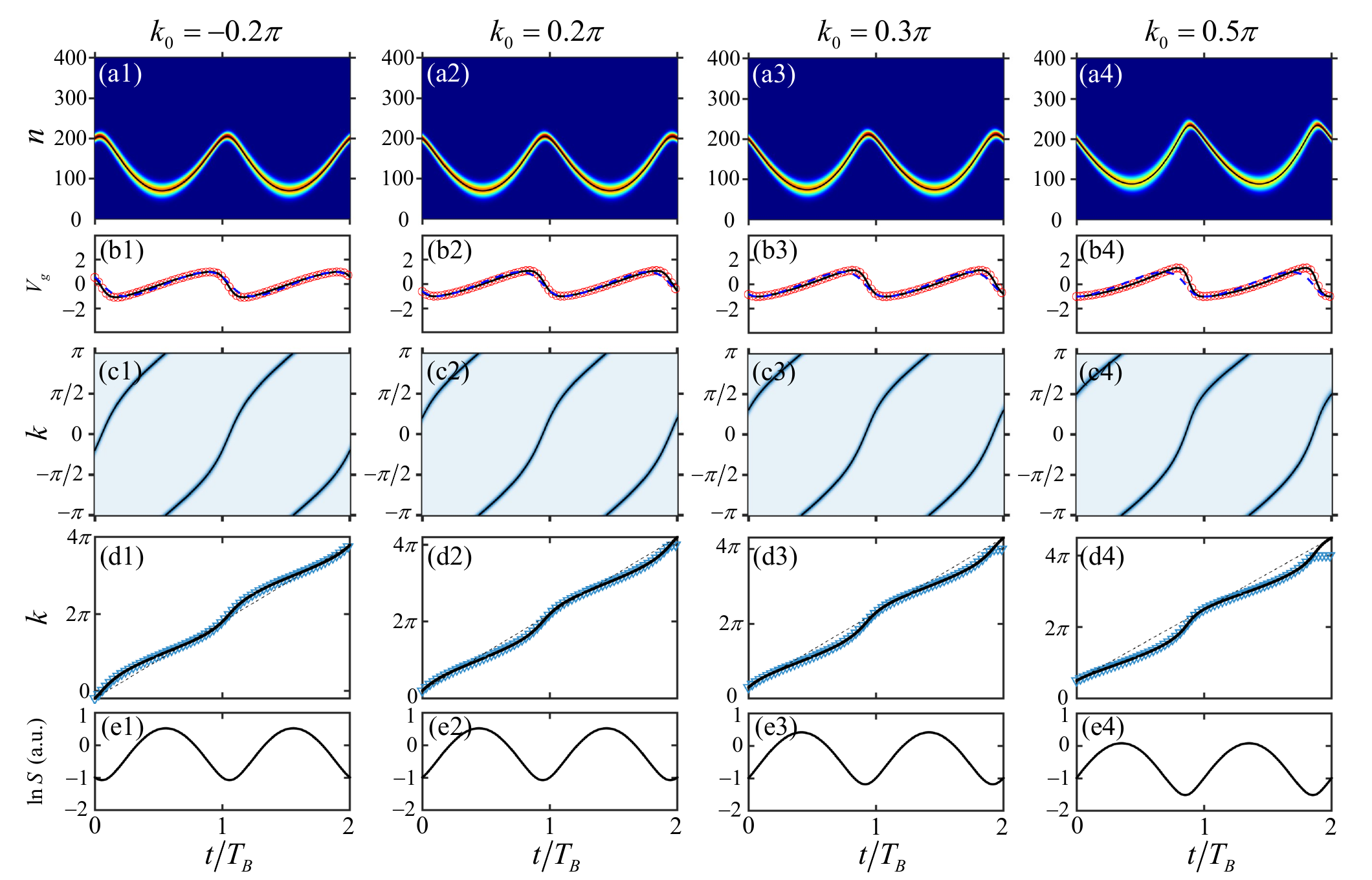}
\caption{Same as those in Fig.~\ref{Fig_infinite_F_sigma_p} with $F_{\text{H}}=-0.015$.}
\label{Fig_infinite_F_k0_n}
\end{figure}

We next discuss the influence of the initial momentum $k_0$. We compare the results for $F_\text{H}>0$ and $F_\text{H}<0$ at $k_0=-0.2\pi,0.2\pi,0.3\pi$, and $0.5\pi$. Based on the transition diagram in Fig.~\ref{Fig_infinite_k0}, for $F_\text{H}>0$, the dynamics for $k_0=\pm0.2\pi$ should be nonsmooth, while the dynamics for  $k_0=0.3\pi$ and  $k_0=0.5\pi$ should be smooth. However, for $F_\text{H}<0$, BOs are always smooth. The simulations in Figs.~\ref{Fig_infinite_F_k0_p} and~\ref{Fig_infinite_F_k0_n} agree with these predictions.

\clearpage

\begin{figure}[h!]
\centering
\includegraphics[width=0.6\linewidth]{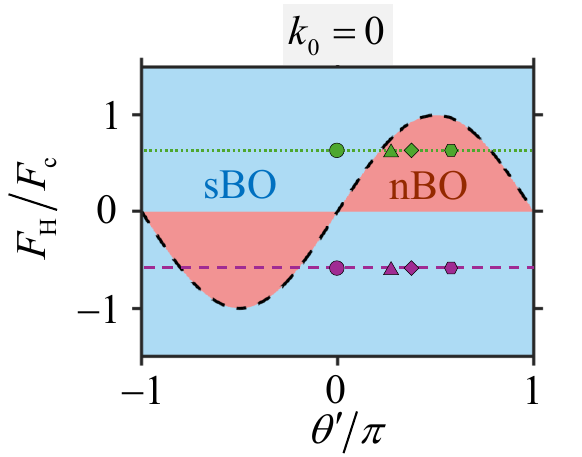}
\caption{Same as Fig.~3(b) in the main text. The green dotted line denote $F_\text{H}=0.015=0.625F_\text{c}$, where the green circle, triangle, diamond, and hexagon denote $\theta'=0$, $0.25\pi$, $0.35\pi$, and $0.65\pi$, respectively. The purple dashed line denote $F_\text{H}=-0.015=-0.625F_\text{c}$, where the purple circle, triangle, diamond, and hexagon denote $\theta'=0$, $0.25\pi$, $0.35\pi$, and $0.65\pi$, respectively.}
\label{Fig_infinite_theta}
\end{figure}

\begin{figure}[h!]
\centering
\includegraphics[width=\linewidth]{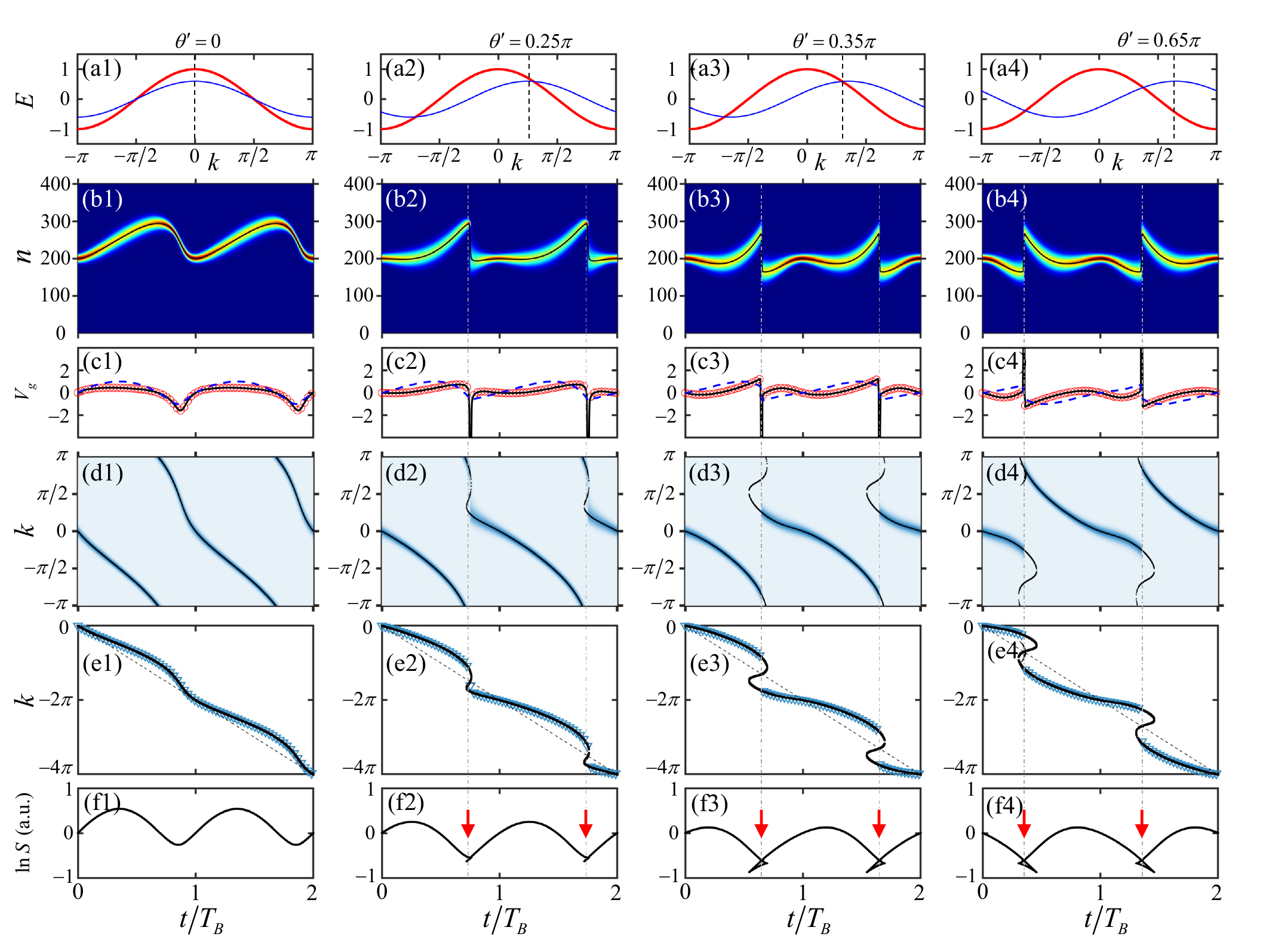}
\caption{Same as those in Fig.~\ref{Fig_infinite_F_sigma_p} with $F_{\text{H}}=0.015$.}
\label{Fig_infinite_F_theta1_p}
\end{figure}

\begin{figure}[h!]
\centering
\includegraphics[width=\linewidth]{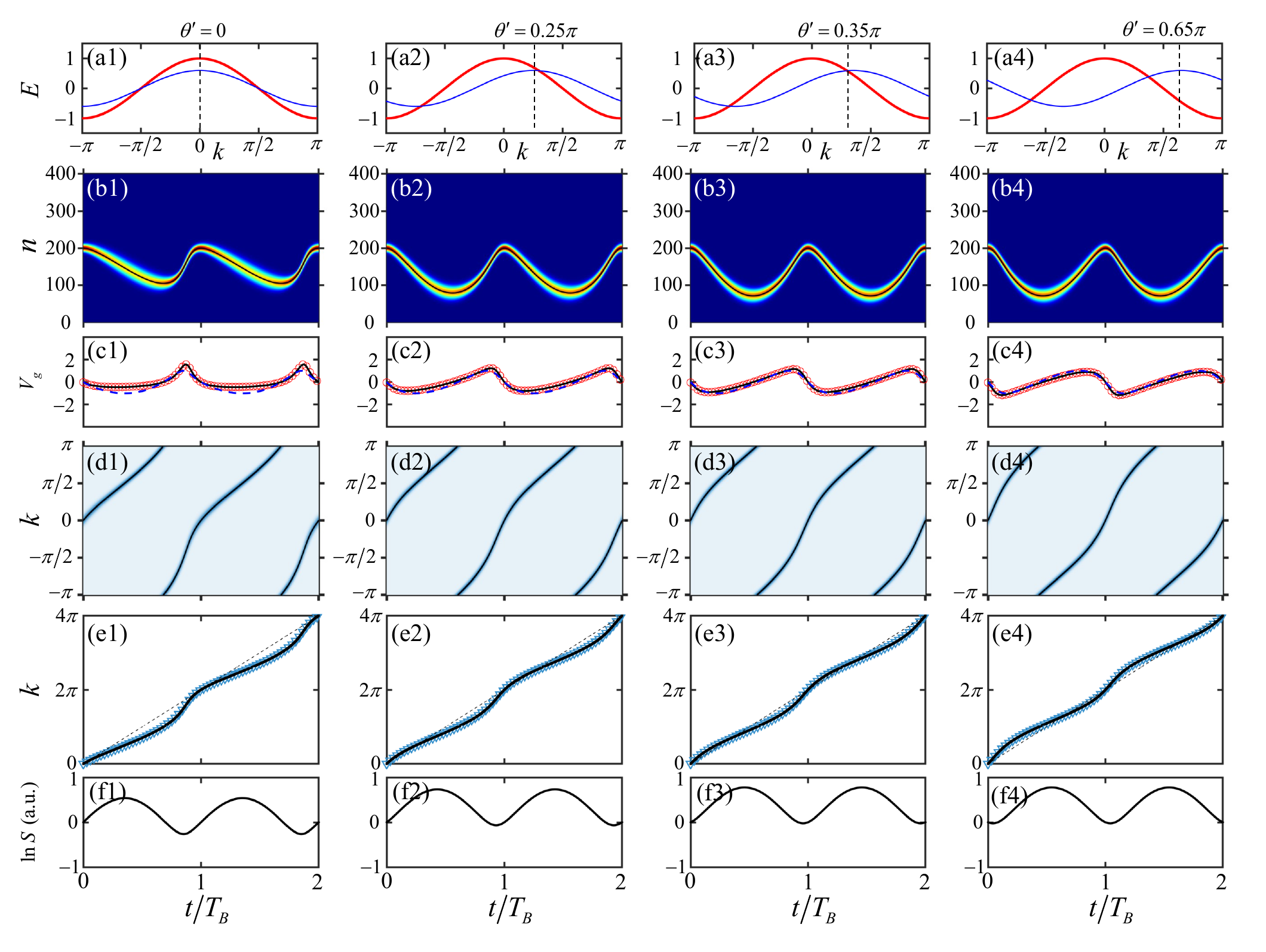}
\caption{Same as those in Fig.~\ref{Fig_infinite_F_sigma_p} with $F_{\text{H}}=-0.015$.}
\label{Fig_infinite_F_theta1_n}
\end{figure}

We last discuss the influence of $\theta'$ on the NH BOs. $\theta'$ shifts the imaginary band, $E_I(k)$, along the $k$ axis, and changes the momentum with the largest $E_I$ to $k=\theta'$. Based on the transition diagram in Fig.~\ref{Fig_infinite_theta}, for $F_\text{H}>0$, the dynamics for $\theta'=0$ should be smooth, while the dynamics for  $\theta'=0.25\pi$ and  $0.35\pi$, and $0.65\pi$ should be nonsmooth. However, for $F_\text{H}<0$, BOs are always smooth. In Fig.~\ref{Fig_infinite_F_theta1_p} and Fig.~\ref{Fig_infinite_F_theta1_n}, we plot the results at $\theta'=0,0.25\pi,0.35\pi$, and $0.65\pi$ for $F_{\text{H}}>0$ and $F_{\text{H}}<0$, respectively. The simulations agree with these predictions. 

Remarkably, we observe that the real-space wave dynamics become discontinuous {with jumps in Figs.~\ref{Fig_infinite_F_theta1_p} (b2-b4)], a feature not observed in previous cases.}
This behavior originates from the asymmetric band structures and the asymmetric $k_{\text{e}}$ near the jump time, as shown in Figs.~\ref{Fig_infinite_F_theta1_p} (a2-a4) and (e2-e4). In these cases, the real-space wave packets corresponding to the dominant momenta before and after the jump are centered at different positions, thereby giving rise to a sudden jump of the wave-packet center. 

\clearpage

\clearpage
\section{VII. NH BOs in unidirectional lattices with OBCs}

In this section, we investigate the NH BOs in lattices with unidirectional couplings under the open boundary conditions, and show that an anomalous phenomenon that the wave can still propagate towards to the direction with vanishing hopping. 
For the wave dynamics in the unidirectional lattice ($J_L=0$) with size $N$, the wave evolution obeys
\begin{align}
    \psi_n(t)=U(t)\psi_n(0)=e^{-iHt}\psi_n(0),
\end{align}
where $H$ is a triangular matrix,
\begin{equation}
H=
\begin{pmatrix}
F_{\text{H}}      & 0      & 0      & \cdots & 0      & 0 \\
J_R    & 2F_{\text{H}}     & 0      & \cdots & 0      & 0 \\
0      & J_R    & 3F_{\text{H}}      & \cdots & 0      & 0 \\
\vdots & \vdots & \vdots & \ddots & \vdots & \vdots \\
0      & 0      & 0      & \cdots & (N-1)F_{\text{H}}  & 0 \\
0      & 0      & 0      & \cdots & J_R    & NF_{\text{H}} 
\end{pmatrix}.
\end{equation}

\begin{figure}[ht!]
\centering
\includegraphics[width=0.8 \linewidth]{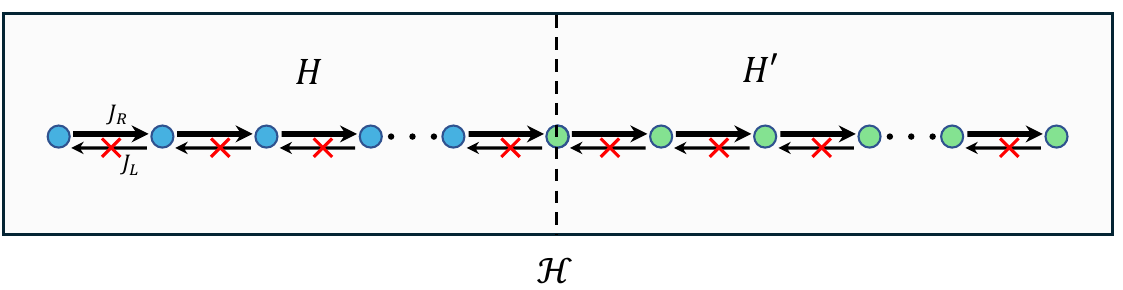}
\caption{Description of wave dynamics without reflection at an unidirectional lattice by constructing another lattice $H'$.}
\label{Fig_OBC}
\end{figure}

We first demonstrate the absence of reflection at the right boundary. We consider a larger lattice which consists of $H$ and another same-size $H'$ at the right, as shown in Fig.~\ref{Fig_OBC}. The Hamiltonian for the new lattice is
\begin{align}
    \mathcal{H}=\begin{pmatrix} H &0\\
    B&H' \end{pmatrix}, \label{H}
\end{align}
where $B$ is the connection matrix between $H$ and $H'$.

The time evolution matrix is then
\begin{align}
    e^{-i\mathcal{H}t}=\sum_{n=0}^{\infty}\frac{(-it)^n}{n!}\mathcal{H}^n
\end{align}
From Eq.~(\ref{H}), we have
\begin{align}
    \mathcal{H}^2=\begin{pmatrix} H &0\\
    B&H' \end{pmatrix}\begin{pmatrix} H &0\\
    B&H' \end{pmatrix}=\begin{pmatrix} H^2 &0\\
    BH+H'B&H'^2\end{pmatrix}.
\end{align}
Therefore $e^{-i\mathcal{H}t}$ is also a triangular matrix, and can be written as 
\begin{align}
    e^{-i\mathcal{H}t}\equiv\begin{pmatrix} U_{11}(t) &0\\
    U_{21}(t)&U_{22}(t) \end{pmatrix}.
\end{align}

From the Schrödinger equation,
\begin{align}
    i\frac{dU}{dt}=\mathcal{H}U,
\end{align}
we have
\begin{align}
    i\dot{U}_{11}={H}U_{11},~i\dot{U}_{22}={H'}U_{22}.
\end{align}
Therefore, we have
\begin{align}
    U_{11}=e^{-iHt},~U_{22}=e^{-iH't},~~i\dot{U}_{21}=Be^{-iHt}+H'U_{21}
\end{align}
Assuming $U_{21}(t)=e^{-iH't}Y(t)$, then we have

\begin{align}
   i\dot{U}_{21}=Be^{-iHt}+H'U_{21}=ie^{-iH't}\dot{Y}(t)+H'e^{-iH't}Y(t),
\end{align}
which gives
\begin{align}
    ie^{-iH't}\dot{Y}(t)=Be^{-iHt}.
\end{align}
Then,
\begin{align}
    Y(t)=-i\int_{0}^{t}e^{iH's}Be^{-iHs}ds,
\end{align}
and 
\begin{align}
    U_{21}(t)=e^{-iH't}Y(t)=-ie^{-iH't}\int_{0}^{t}e^{iH's}Be^{-iHs}ds=-i\int_0^te^{-iH'(t-s)}Be^{-iHs}ds.
\end{align}

Therefore the time-evolution matrix is\begin{align}
    e^{-i\mathcal{H}t}=\begin{pmatrix} e^{-iHt} &0\\
    -i\int_0^te^{-iH'(t-s)}Be^{-iHs}ds&e^{-iH't}\end{pmatrix}.
\end{align}
This equation shows that the wave evolution {inside} the lattice $H$ is independent of the specific form of $H'$. Therefore, 
the wave dynamics {inside} the finite lattice $H$ is same as that in the larger-size lattice, where the boundary effect can be ignored. 

\begin{figure}[h!]
\centering
\includegraphics[width=0.7 \linewidth]{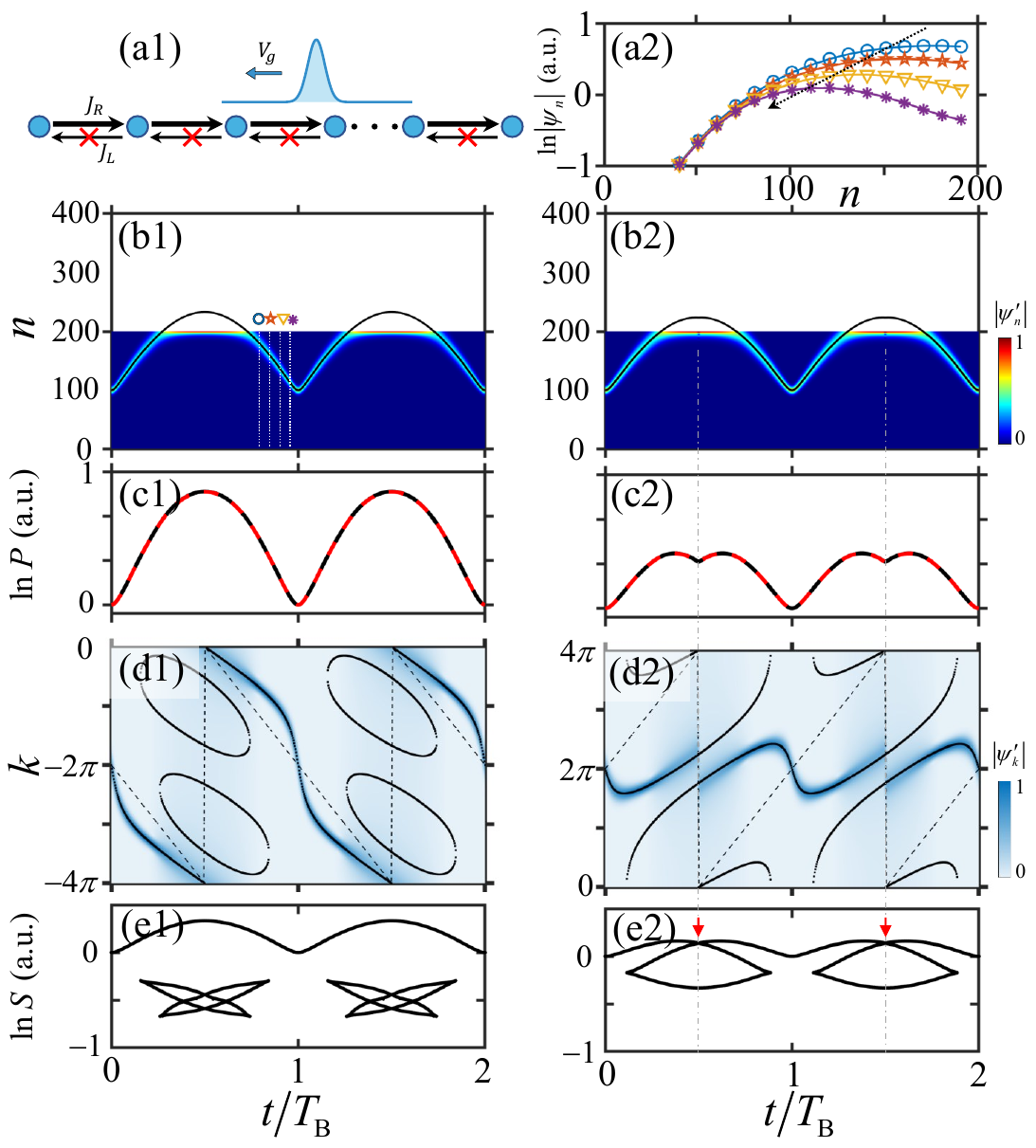}
\caption{(a1) Anomalous wave propagation toward the direction with vanishing hopping. (a2) Wave-packet amplitudes at different times indicated by the markers in (b1), where the black dotted line indicates the leftward shift of the wave-packet center. (b1,b2) Real-space wave-packet dynamics for $N=200$, where the black solid lines show the predicted $\langle n(t)\rangle$ obtained from Eq.~(\ref{n}). (c1,c2) Time evolution of the total wave intensity $P(t)$. (d1,d2) and (e1,e2) Same as those in Fig.~\ref{fig3_SP}. $\sigma=3,J_L=0$, and $J_R=1$. }
\label{fig_uni}
\end{figure}

We investigate NH BOs in the finite lattice with unidirectional hopping under OBCs, where $J_L=0,J_R\ne0$ [Fig.~\ref{fig_uni}(a1)]. The corresponding energy band is $E(k)=J_Re^{-ik}$, which forms a circle in the complex energy plane and exhibits a \textit{circular point-gap topology}. Since the wave dynamics can be affected only by the left side of the lattice, no reflection occurs at the right boundary. Consequently, the wave dynamics is identical to that of the infinite system during the range of the finite lattice. This is demonstrated in Figs.~\ref{fig_uni}(b1) and~\ref{fig_uni}(b2), where the center of mass trajectory predicted by Eq.~(\ref{n}) matches well with the numerical wave dynamics. Meanwhile, the total wave intensity, $P(t)=\sum_{n=1}^{N}|\psi_n(t)|^2$, shown by the black solid lines, coincides with its infinite-lattice counterpart during the range of the finite lattice (red dashed lines) [Figs.~\ref{fig_uni}(c1) and~\ref{fig_uni}(c2)]. Owing to this equivalence, the wave dynamics under OBCs can be understood in terms of the corresponding infinite-lattice dynamics. The wave packet passes through the right boundary, leaves behind a boundary-localized tail, and subsequently reenters the lattice, which hence induces \textit{periodic temporal Goos–Hänchen shifts} at the lattice boundary [Figs.~\ref{fig_uni}(b1) and~\ref{fig_uni}(b2)]. In addition, the corresponding momentum-space evolution also follows the predicted trajectory $k_{\text{e}}(t)$ in Eq.~(\ref{ke_SM}) [Figs.~\ref{fig_uni}(d1) and~\ref{fig_uni}(d2)] with the largest amplitude $|S(k_{\text{e}},t)| $[Figs.~\ref{fig_uni}(e1) and~\ref{fig_uni}(e2)]. 

Moreover, a counterintuitive leftward wave propagation is observed for $J_L=0$ [e.g., Fig.~\ref{fig4}(b1)]. This anomalous behavior arises from \textit{amplitude redistribution} rather than a physical "transport" of the wave packet, as a consequence of the breakdown of energy conservation in NH systems. Fig.~\ref{fig4}(a2) plots the amplitudes of the wave packet at four times before $t=T_\text{B}$, corresponding to  the leftward propagation indicated by the different markers in Fig.~\ref{fig4}(b1). We observe that the right part of the wave-packet amplitude decays rapidly, whereas the left part reduces slowly, resulting in a leftward shift of the center of mass, as indicated by the black dotted line.

\section{VIII. Analytical analysis on the zero group velocity for single-site excitation}

In the section "Single-site excitation" of the main text, we reveal an anomalous wave phenomenon in which the momentum evolves continuously while the group velocity remains zero, as shown in Figs.~4(b2-d2) of the main text. Here we analyze analytically the physical reason behind this phenomenon. 

The evolution of the center of mass is
    \begin{align}
    \langle n(t) \rangle&=n_0+\left\langle\frac{E_R(k+F_{\text{H}}t)-E_R(k)}{F_{\text{H}}}\right\rangle.\nonumber\\
    &=n_0+\frac{2\mathcal{J}}{F_{\text{H}}} \left[\left\langle \cos(k+F_{\text{H}}t)- \cos(k)\right\rangle\right],
\end{align}
where we have used $E_R=2\mathcal{J}\cos(k-\theta)$ with $\theta=0$. 
Since the momentum distribution $|S(k,t)|=e^{M(k,t)}=e^{\Phi_I(k,t)}=e^{(4\mathcal{J}'/F_{\text{H}})\sin(F_{\text{H}}t/2)\cos(k+F_{\text{H}}t/2-\theta')}$, we then have
\begin{align}
    \langle n(t) \rangle= n_0+\frac{2\mathcal{J}}{F_{\text{H}}}\frac{\int_{-\pi}^{\pi}dk[\cos(k+F_{\text{H}}t)- \cos(k)]|S(k,t)|^2}{\int_{-\pi}^{\pi}dk|S(k,t)|^2}\equiv n_0+\frac{2\mathcal{J}}{F_{\text{H}}}\frac{P}{D}.
\end{align}
Here
\begin{align}
    P&=\int_{-\pi}^{\pi}dk[\cos(k+F_{\text{H}}t)- \cos(k)]|S(k,t)|^2=-2\sin{\frac{F_{\text{H}}t}{2}}\int_{-\pi}^{\pi}dk\sin\left(k+\frac{F_{\text{H}}t}{2}\right)e^{V\cos(k+F_{\text{H}}t/2-\theta')}, \nonumber\\
    D&=\int_{-\pi}^{\pi}dk|S(k,t)|^2=\int_{-\pi}^{\pi}dke^{V\cos(k+F_{\text{H}}t/2-\theta')},
\end{align}
with $V=(8\mathcal{J}'/F_{\text{H}})\sin(F_{\text{H}}t/2)$.

Let $k'=k+F_{\text{H}}t/2-\theta'$, we then have
\begin{align}
    P&=-2\sin{\frac{F_{\text{H}}t}{2}}\int_{-\pi+F_{\text{H}}t/2-\theta'}^{\pi+F_{\text{H}}t/2-\theta'}dk'\sin\left(k'+\theta'\right)e^{V\cos(k')}\nonumber\\
    &=-2\sin{\frac{F_{\text{H}}t}{2}}\int_{-\pi}^{\pi}dk'\sin\left(k'+\theta'\right)e^{V\cos(k')},
\end{align}
where the change of integral range is due to the periodic function $\sin\left(k'+\theta'\right)e^{V\cos(k')}$ with a period of $2\pi$. 

Since 
\begin{align}
    \sin(k'+\theta')=\sin(k')\cos(\theta')+\cos(k')\sin(\theta'),
\end{align}
we then obtain 
\begin{align}
    P&=-2\sin{\frac{F_{\text{H}}t}{2}}\int_{-\pi}^{\pi}dk'\sin\left(k'+\theta'\right)e^{V\cos(k')}\nonumber\\
    &=-2\sin{\frac{F_{\text{H}}t}{2}}\left[ \cos(\theta') \int_{-\pi}^{\pi}dk'\sin\left(k'\right)e^{V\cos(k')}+\sin(\theta') \int_{-\pi}^{\pi}dk'\cos\left(k'\right)e^{V\cos(k')}\right]\nonumber\\
    &=-2\sin{\frac{F_{\text{H}}t}{2}}\sin(\theta') \int_{-\pi}^{\pi}dk'\cos\left(k'\right)e^{V\cos(k')},
\end{align}
where the first term is zero, as $\sin(k')$ is an odd function and $e^{V\cos(k')}$ is an even function.

In the same way, we obtain 
\begin{align}
    D=\int_{-\pi}^{\pi}dke^{V\cos(k+F_{\text{H}}t/2-\theta')}=\int_{-\pi}^{\pi}dk'e^{V\cos(k')}.
\end{align}
Therefore, 
\begin{align}
    \frac{P}{D}=-2\sin{\frac{F_{\text{H}}t}{2}}\sin(\theta')\frac{ \int_{-\pi}^{\pi}dk'\cos\left(k'\right)e^{V\cos(k')}}{\int_{-\pi}^{\pi}dk'e^{V\cos(k')}}.
\end{align}
We use the modified Bessel function of the first kind \cite{abramowitz1972handbook},
\begin{align}
    I_n(V)=\frac{1}{2\pi}\int_{-\pi}^{\pi}e^{V\cos(k')}e^{-ink'}dk'.
\end{align}
For $n=0$, we have
\begin{align}
    I_0(V)=\frac{1}{2\pi}\int_{-\pi}^{\pi}e^{V\cos(k')}dk',
\end{align}
which gives
\begin{align}
    \int_{-\pi}^{\pi}e^{V\cos(k')}dk'=2\pi I_0(V).
\end{align}
For $n=1$, we have
\begin{align}
    I_1(V)=\frac{1}{2\pi}\int_{-\pi}^{\pi}e^{V\cos(k')}e^{-ik'}dk'=\frac{1}{2\pi}\int_{-\pi}^{\pi}\left[\cos(k')-i\sin(k')\right]e^{V\cos(k')}=\frac{1}{2\pi}\int_{-\pi}^{\pi}\cos(k')e^{V\cos(k')},
\end{align}
which gives
\begin{align}
    \int_{-\pi}^{\pi}\cos(k')e^{V\cos(k')}=2\pi I_1(V).
\end{align}
Therefore,
\begin{align}
    \frac{P}{D}=-2\sin{\frac{F_{\text{H}}t}{2}}\sin(\theta')\frac{ \int_{-\pi}^{\pi}dk'\cos\left(k'\right)e^{V\cos(k')}}{\int_{-\pi}^{\pi}dk'e^{V\cos(k')}}=-2\sin{\frac{F_{\text{H}}t}{2}}\sin(\theta')\frac{I_1(V)}{I_0(V)},
\end{align}
which hence gives a analytical expression of the center of mass
\begin{align}
    \langle n(t) \rangle= n_0+\frac{2\mathcal{J}}{F_{\text{H}}}\frac{P}{D}=n_0-\frac{4\mathcal{J}}{F_{\text{H}}}\sin{\frac{F_{\text{H}}t}{2}}\sin(\theta')\frac{I_1(V)}{I_0(V)},
\end{align}
with $V=(8\mathcal{J}'/F_{\text{H}})\sin(F_{\text{H}}t/2)$.

When $\theta'=0$, the center of mass $\langle n(t) \rangle= n_0$, and hence the corresponding group velocity $V_A=d\langle n(t) \rangle/dt=0$.

In the numerical simulations of this work, we have used the arbitrary precision library Advanpix in Matlab to improve the numerical precision \cite{Advanpix}.




\end{document}